\documentclass{isprs}
\usepackage[T1]{fontenc}
\usepackage[justification=justified]{caption} 
\usepackage{subcaption}
\usepackage{setspace}
\usepackage{geometry} % added 27-02-2014 Markus Englich
\usepackage{epstopdf}
\usepackage{natbib}
\usepackage[labelsep=period]{caption}  % added 14-04-2016 Markus Englich - Recommendation by Sebastian Brocks

\usepackage{color}
\usepackage{graphicx}
\usepackage[dvipsnames]{xcolor}
\usepackage{tikz}
\usetikzlibrary{arrows,patterns}
\usepackage{verbatim}
\usepackage{amssymb, amsmath} 
\usepackage{pgfplots}
\usepackage{rotating}
\usepackage{url}

\usepackage{array}
\newcolumntype{L}[1]{>{\raggedright\let\newline\\\arraybackslash\hspace{0pt}}m{#1}}
\newcolumntype{C}[1]{>{\centering\let\newline\\\arraybackslash\hspace{0pt}}m{#1}}
\newcolumntype{R}[1]{>{\raggedleft\let\newline\\\arraybackslash\hspace{0pt}}m{#1}}
\newcolumntype{M}{>{\centering\arraybackslash} m{3cm} }

\begin{document}

\title{
Sensor-Topology based simplicial complex reconstruction from mobile laser scanning}

% KAO: Remove extra spacing
%%% Penser à enlever avant la soumission
\author{
St\'ephane Guinard\textsuperscript{$\ast$}, Bruno Vallet\textsuperscript{$\ast$}}

% KAO: Remove extra newline
%%% Penser à enlever avant la soumission
\address{
	\textsuperscript{$\ast$}Universit\'e Paris-Est, LASTIG MATIS, IGN, ENSG, \\
	73 avenue de Paris, 94160 Saint-Mand\'e, France \\
	 (stephane.guinard, bruno.vallet)@ign.fr
}

% If the corresponding author is NOT the final author, always add a % space before the subsequent comma, i.e.
% first author name\textsuperscript{a,}\thanks{Corresponding author} , % second author name \textsuperscript{b}, etc.
% thanks to Niclas Borlin 05-05-2016

\commission{II, }{ } %This field is optional.
\workinggroup{II/4} %This field is optional.
\icwg{}   %This field is optional.

\abstract{
We propose a new method for the reconstruction of simplicial complexes (combining points, edges and triangles) from 3D point clouds from Mobile Laser Scanning (MLS). {Our main goal is to produce a reconstruction of a scene that is adapted to the local geometry of objects 
%(e.g. poles are linear and usually correspond to a few aligned points in the scan
.} Our method uses the inherent topology of the MLS sensor to define a spatial adjacency relationship between points. We then investigate each possible connexion between adjacent points and filter them by searching collinear structures in the scene, or structures perpendicular to the laser beams. Next, we create triangles for each triplet of self-connected edges. Last, we improve this method with a regularization based on the co-planarity of triangles and collinearity of remaining edges. We compare our results to a naive simplicial complexes reconstruction based on edge length.
}


\maketitle

\section{Introduction}

LiDAR scanning technologies have become a widespread and direct mean for acquiring a precise sampling of the geometry of scenes of interest. However, unlike images, LiDAR point clouds do not always have a natural topology (4- or 8-neighborhoods for images) allowing to recover the continuous nature of the acquired scenes from the individual samples. This is why a large amount of research work has been dedicated into recovering a continuous surface from a cloud of point samples, which is a central problem in geometry processing. Surface reconstruction generally aims at reconstructing triangulated surface meshes from point clouds, as they are the most common numerical representation for surfaces in 3D, thus well adapted for further processing. Surface mesh reconstruction has numerous applications in various domains:
\begin{itemize}
\item Visualization: a surface mesh is much more adapted to visualization than a point cloud, as the visible surface is interpolated between points, allowing for a continuous representation of the real surface, and enabling the estimation of occlusions, thus to render only the visible parts of the scene.
\item Estimation of differential quantities such as surface normals and curvatures.
\item Texturing: a surface mesh can be textured (images applied on it) allowing for photo-realistic rendering. In particular, when multiple images of the acquired scene exists, texturing allows to fusion and blend them all into a single 3D representation.
\item Shape and object detection and reconstruction: these high level processes benefit from surface reconstruction since it solves the basic geometric ambiguity (which points are connected by a real surface in the real scene ? ).
\end{itemize}

\begin{figure*}[t]
\centering
\includegraphics[width=1\textwidth]{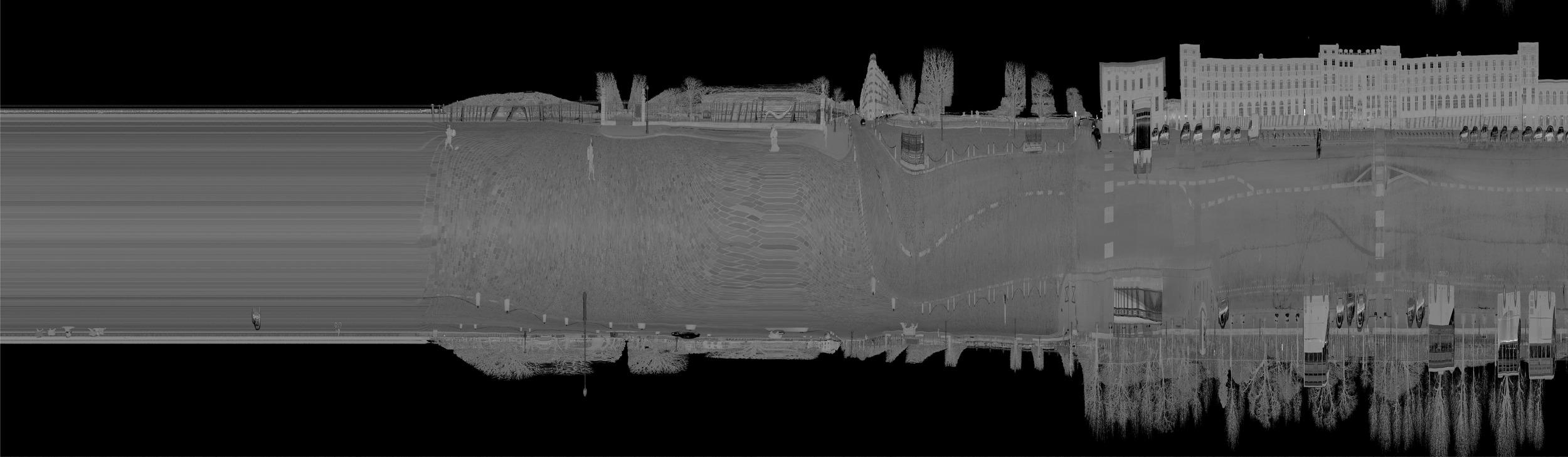}
\caption{Echo intensity of a MLS displayed in sensor topology: vertical axis is the angle $\theta$, horizontal axis is the line number, equivalent to time as the scanner acquires a constant number of lines per second. Horizontal resolution depends on vehicle speed (the left part is constant because the vehicle is stopped).}
\label{fig:sensorTopo}
\end{figure*}

\begin{figure*}[t]
\centering
\includegraphics[width=1\textwidth]{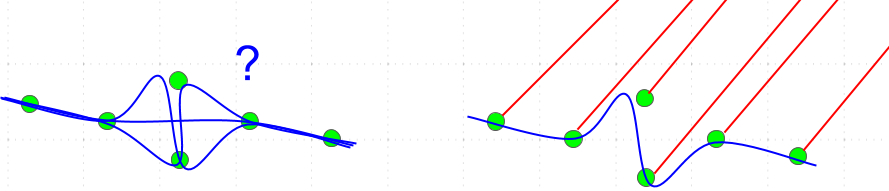}
\caption{Left: a 2D point cloud (green) and possible reconstructions (blue). Right: knowing the LiDAR rays allows to solve the ambiguity}
\label{fig:ambig}
\end{figure*}

In practice, existing surface reconstruction algorithms often consider that their input is a set of $(x,y,z)$ coordinates, possibly with normals. However, most LiDAR scanning technologies provide more than that: the sensors have a logic of acquisition that provides a sensor topology \citep{xiao2013change,vallet2015terramobilita}. For instance, planar scanners acquire points along a line that advances with the platform (plane, car, ...) it is mounted on. Thus each point can be naturally connected to the one before and after him along the line, and to its equivalent in the previous and next lines (see Figure \ref{fig:sensorTopo}).
% BV: trop précis pour une intro: If the number of pulses per line is not integer, the point can be connected to the two closest points (in angle) from the previous and next lines, leading to a 6-neighborhood and an hexagonal topology.
Fixed LiDARs scan in spherical $(\theta, \phi)$ coordinates which also imply a natural connection of each point to the previous and next along these two angles. Some scanner manufacturers exploit this topology by proposing visualization and processing tools in 2.5D (depth images in $(\theta, \phi)$) rather than 3D. Moreover, LiDAR scanning can provide a meaningful information that is the position of the LiDAR sensor for each points, resulting in a ray along which we are sure that space is empty. This information can also disambiguate surface reconstruction as illustrated in Figure \ref{fig:ambig}. This is why we decided to investigate the use of the sensor topology inherent to a MLS, to perform a 3D reconstruction of a point cloud.

\begin{figure}[t]
\centering
\includegraphics[width=0.6\columnwidth]{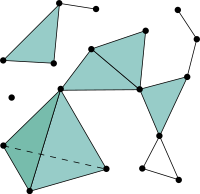}
\caption{A simplicial complex consists of simplices of dimension 0 (points), 1 (edges) and 2 (triangles) (source Wikipedia)}
\label{fig:simplicialComplexe}
\end{figure}

%Moreover we argue that, due to the acquisition density and scaling, it may not be possible to interpret some structures as 3D objects. For instance, poles or electric wires are most of the time represented in the cloud by a single line of points. Therefore, we chose to represent our final structure as a simplicial complex, containing 2D simplexes (triangles) for planar structures (roads or facades), 1D simplexes (segments) for linear structures (poles, wires) and 0D simplexes (points) for areas where, according to the Nyquist - Shannon sampling theorem we don't have enough information to interpret the cloud. We remark that this last case will be useful for noisy areas, like foliage of trees where each echo can land on separate leaves or twigs and we cannot design an appropriate representation of the point cloud.

Secondly, the geometry processing community has mainly focused on reconstruction of rather smooth objects, possibly with sharp edges, but with a sampling density sufficient to consider that the object is a 2-manifold, which means that it is locally 2-dimensional. Thus these methods do not extend well to real scenes where such a guarantee is hardly possible. In particular, scans including poles, power lines, wires, ... almost never allow to create triangles on these structures because their widths (a few mm to a few cm) is much smaller than the scanning resolution. Scans of highly detailed structures (such as tree foliage for instance) even have a 0-dimensional nature: individual points should not even be connected to any of their neighbors. Applying the Nyquist-Shannon theorem to the the range in sensor space tells us that if the geometric frequency (frequency of the range signal in sensor space) is higher than half the sampling frequency (frequency of the samples in sensor space), then some (geometric) signal will be lost, which happens in the cases stated above for instance. Because of this, we should aim at reconstructing triangles only when the Shannon condition is met in the two dimensions, but only edges when the geometric frequency is too high in 1 dimension and points when the geometric frequency is too high in the 2 dimensions. Triangles, edges and points are called simplices, which are characterized by their dimension $d$ ($0=$ vertices, $1=$ edges, $2=$ triangles). If we add the constraint that edges can only meet at a vertex and triangles can only meet at an edge or vertex, the resulting mathematical object is called a \textit{simplicial complex} as illustrated in Figure \ref{fig:simplicialComplexe}. The aim of this paper is to propose a method to reconstruct such simplicial complexes from a LiDAR scan.

\section{State of the art}

3D surface mesh reconstruction from point clouds has been a major issue in geometry processing for the last decades.
3D reconstruction can be performed from oriented \citep{kazhdan2013screened} or unoriented point sets \citep{alliez2007voronoi}. Data itself can come from various sources: Terrestrial Laser Scanning \citep{pu2009knowledge}, Aerial Laser Scanning \citep{dorninger2008comprehensive} or Mobile Laser Scanning. We refer the reader to  \citet{berger2014state} for a general review of surface reconstruction methodologies and focus our state of the art on surface reconstruction from Mobile Laser Scanning (MLS) and on simplicial complexes reconstruction, which are the two specificities of our approach.

MLS have been used for the past years mostly for the modeling of outdoors environments, usually in urban scenes. \citet{becker2009grammar} propose an automatically-generated grammar for the reconstruction buildings, whereas \citet{rutzinger2010detection} focus more specifically on tree shapes reconstruction. MLS has also been useful for specific indoor environments: \citet{zlot2014efficient} used a MLS in an underground mine to obtain a 3D model of the tunnels.

The utility of simplicial complexes for the reconstruction of 3D point clouds has been expressed by \citet{popovic1997progressive} as a generalization manner to simplify 3D meshes. Simplicial complexes are also used to simplify defect-laden point sets as a way to be robust to noise and outliers using optimal transport \citep{de2011optimal,digne2014feature} or alpha-shapes \citep{bernardini1997sampling}.

As explained in introduction, the aim of this paper is to propose a reconstruction method that combines two advantages:
\begin{enumerate}
\item Reconstruction of a simplicial complexe instead of a surface mesh, adapting the local dimension to that of the local structure.
\item Exploiting the sensor topology both to solve ambiguities and to speed up computations.
\end{enumerate}
The two objectives are tackled at once by proposing a new criteria to define which simplices from the sensor topology should belong to the reconstructed simplicial complex.

\section{Methodology}

As explained above, sensor topology yields in general a regular mesh structure with a 6-neighborhood that can be used to perform a surface mesh reconstruction. This reconstruction is however very poor as all depth discontinuities will be meshed, so very elongated triangles will be constructed between objects and their background. This section investigates criteria to remove these triangles, while possibly keeping some of their edges. As all input points will be kept, the resulting reconstruction combines points and edges and triangles based on these points, which is called a \textit{simplicial complex} in mathematics.

\subsection{Objectives}

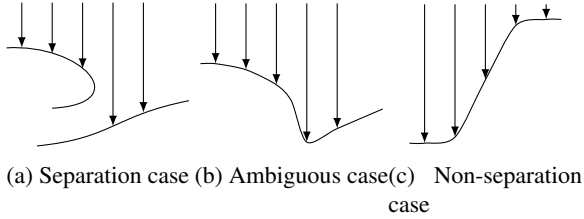
\begin{figure}[t]
\centering
%\begin{subfigure}[t]{.3\textwidth}
\begin{subfigure}[t]{.15\textwidth}
\centering
  \begin{tikzpicture}[scale=0.2]
    \draw[-latex,black] (0,5) -- (0,2);
    \draw[-latex,black] (2,5) -- (2,1.6);
    \draw[-latex,black] (4,5) -- (4,0.6);
    \draw[-latex,black] (6,5) -- (6,-3.2);
    \draw[-latex,black] (8,5) -- (8,-2.3);
    \draw (-1,2) ..controls +(5,0.3) and +(5.5,0.3).. (2,-2); % arc 1
    \draw (11,-1.5) ..controls (6,-2.5) and +(5,0.5).. (1,-4.5); % arc 2
  \end{tikzpicture}
  \caption{Separation case}
  \label{fig:sep}
  \end{subfigure}%
  \begin{subfigure}[t]{.15\textwidth} 
  \centering
  \begin{tikzpicture}[scale=0.2]
    \draw[-latex,black] (0,6) -- (0,2);
    \draw[-latex,black] (2,6) -- (2,1.6);
    \draw[-latex,black] (4,6) -- (4,0.6);
    \draw[-latex,black] (6,6) -- (6,-3.2);
    \draw[-latex,black] (8,6) -- (8,-2.3);
    \draw plot [smooth] coordinates {(-1,2) (0,2) (2,1.6) (4,.6) (5,-.5) (6,-3.2) (8,-2.3) (11,-1)};
  \end{tikzpicture}
  \caption{Ambiguous case}
  \label{fig:ambigu}
  \end{subfigure}%
  %\caption{Separation case}
  %\label{fig:sep}
  %\end{subfigure}
  \begin{subfigure}[t]{0.15\textwidth}
  \centering
  \begin{tikzpicture}[scale=0.2]
    \draw[-latex,black] (10,4) -- (10,3);
    \draw[-latex,black] (2,4) -- (2,-5.2);
    \draw[-latex,black] (4,4) -- (4,-4.8);
    \draw[-latex,black] (6,4) -- (6,-1);
    \draw[-latex,black] (8,4) -- (8,2.6);
    \draw plot [smooth] coordinates {(1,-5.2) (2,-5.2) (4,-4.8) (6,-1) (8,2.6) (10,3) (11,3)}; % moins smooth que dessous mais plus facile pour faire une courbe
    %\draw (11,3) ..controls (7,4.5) and +(6,-1.5).. (1,-5);
  \end{tikzpicture}
  \caption{Non-separation case}
  \label{fig:no-sep}
  \end{subfigure}
\caption{Illustration of the two cases of important depth difference. The arrows represent the laser beams. Figures \ref{fig:sep} and \ref{fig:ambigu} show the cases where two neighboring echoes have a huge depth difference. They can either fall on two different objects or on a same object and we have no hint to distinguish these two cases.
%Figure \ref{fig:sep} shows the case where two neighboring echoes have a huge depth difference. In this case we can't find out if the echoes fell on different objects or not. We decide that we won't connect these echoes. 
Figure \ref{fig:no-sep} shows the case where three or more echoes are approximately aligned, with a huge depth difference. In this case we want to reconstruct edges between these echoes because it may correspond to a grazing surface.}
\label{fig:sep-cases}
\end{figure}

Our main objective is to determine which adjacent points (in sensor topology) should be connected to form edges and triangles. We consider that we may be facing a discontinuity when the depth difference between two neighboring echoes is high. This depth difference is computed from a sensor viewpoint, which implies that a large depth difference may correspond to two cases: either the echoes fell on two different objects with a notable depth difference%(\ref{fig:sep})
, or they fell on a grazing surface (nearly parallel to the laser beams direction) as shown in figure \ref{fig:sep-cases}. When two neighboring echoes have a large depth difference, there is no way we can guess whether they are located on two separate objects (\ref{fig:sep}) or a grazing surface (\ref{fig:ambigu}). The core idea of our filtering is that the only hint we can rely on to distinguish between these two cases is that if at least three echoes with a large depth differences are aligned (\ref{fig:no-sep}), we probably are in the grazing surface case rather than on separate objects.
% BV: redondant We will distinguish two different cases: the first case is when we are facing two neighboring echoes with a huge depth difference as shown in figure (\ref{fig:sep}). We will assume that this case occurs when the echoes fell on two different objects with a notable depth difference. The second case occurs when three or more neighboring echoes with a huge depth difference are approximately aligned. This last case may correspond to a grazing surface from the sensor viewpoint (i.e. when the laser is scanning an object perpendicular to its movement) like in figure \ref{fig:no-sep}.

To perform our reconstruction, we consider each echo as an independent point. First, we define a neighborhood relationship between echoes in the sensor topology. Then, we create edges, based on the echoes and add triangles based on the edges. Last, we regularize the computed simplicial complex according to the local geometric consistency of the retrieved simplexes.

\subsection{Neighborhood in sensor topology}

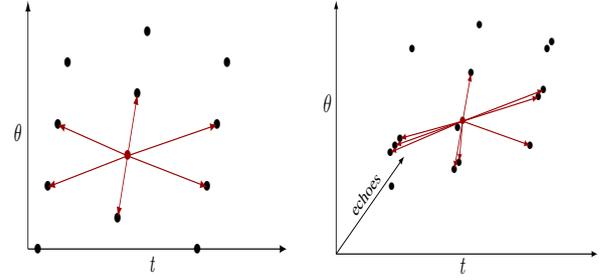
\begin{figure}[t]
\centering
\begin{subfigure}[t]{0.23\textwidth}
\centering
\resizebox{0.95\textwidth}{0.95\textwidth}{
\begin{tikzpicture}
\draw[-latex] (0,0) -- (0,3.2);
\draw[-latex] (0,0) -- (5.2,0);
\node at (-0.2,1.5) {$\theta$};
\node at (2.5,-0.2) {$\mathit{t}$};

\node at (0.2,0) {$\bullet$};
\node at (0.4,0.8) {$\bullet$};
\node at (0.6,1.6) {$\bullet$};
\node at (0.8,2.4) {$\bullet$};
\node at (1.8,0.4) {$\bullet$};
\node at (2,1.2) {\textcolor{red!60!black}{$\bullet$}};
\node at (2.2,2) {$\bullet$};
\node at (2.4,2.8) {$\bullet$};
\node at (3.4,0) {$\bullet$};
\node at (3.6,0.8) {$\bullet$};
\node at (3.8,1.6) {$\bullet$};
\node at (4,2.4) {$\bullet$};

\draw[-latex,red!60!black] (2,1.2) -- (0.4,0.8);
\draw[-latex,red!60!black] (2,1.2) -- (0.6,1.6);
\draw[-latex,red!60!black] (2,1.2) -- (1.8,0.4);
\draw[-latex,red!60!black] (2,1.2) -- (2.2,2);
\draw[-latex,red!60!black] (2,1.2) -- (3.6,0.8);
\draw[-latex,red!60!black] (2,1.2) -- (3.8,1.6);
\end{tikzpicture}}
\caption{The pulse sensor topology forms a 6-neighborhood}
\label{fig:6v}
\end{subfigure}%
\hspace{.1cm}
\begin{subfigure}[t]{0.23\textwidth}
\centering
%\scalebox{1.3}{
\resizebox{0.95\textwidth}{0.95\textwidth}{
\begin{tikzpicture}
\draw[-latex] (-1,-1,0) -- (-1,3.2,0);
\draw[-latex] (-1,-1,0) -- (5.2,-1,0);
\draw[-latex] (-1,-1,0) -- (-1,-1,-4.2);
\node at (-1.2,1.5,0) {\large{$\theta$}};
\node at (2,-1.2,0) {\large{$\mathit{t}$}};
\node[rotate=45] at (-.3,0,0) {{echoes}};

\node at (0.2,0,-.3) {$\bullet$};

\node at (0.4,0.8,0) {$\bullet$};
\node at (0.4,0.8,-.3) {$\bullet$};
\node at (0.4,0.8,.3) {$\bullet$};

\node at (0.8,2.4,0) {$\bullet$};

\node at (1.8,0.4,0) {$\bullet$};
\node at (1.8,0.4,-.3) {$\bullet$};

\node at (2,1.2,0) {\textcolor{red!60!black}{$\bullet$}};
\node at (2,1.2,.3) {$\bullet$};

\node at (2.2,2,0) {$\bullet$};

\node at (2.4,2.8,0) {$\bullet$};

\node at (3.6,0.8,0) {$\bullet$};

\node at (3.8,1.6,0) {$\bullet$};
\node at (3.8,1.6,-.3) {$\bullet$};

\node at (4,2.4,0) {$\bullet$};
\node at (4,2.4,-.3) {$\bullet$};

\draw[-latex,red!60!black] (2,1.2,0) -- (0.4,0.8,0);
\draw[-latex,red!60!black] (2,1.2,0) -- (0.4,0.8,.3);
\draw[-latex,red!60!black] (2,1.2,0) -- (0.4,0.8,-.3);
\draw[-latex,red!60!black] (2,1.2,0) -- (1.8,0.4,0);
\draw[-latex,red!60!black] (2,1.2,0) -- (1.8,0.4,-.3);
\draw[-latex,red!60!black] (2,1.2,0) -- (2.2,2,0);
\draw[-latex,red!60!black] (2,1.2,0) -- (3.6,0.8,0);
\draw[-latex,red!60!black] (2,1.2,0) -- (3.8,1.6,0);
\draw[-latex,red!60!black] (2,1.2,0) -- (3.8,1.6,-.3);
\end{tikzpicture}}
\caption{Echo sensor topology: each echo is connected to all echoes of all neighboring pulses}
\label{fig:multi-echo}
\end{subfigure}
\caption{Definition of neighborhood in sensor space. For each figure, the points considered is colored in red, and connection is denoted by a red arrow.}
\end{figure}

The sensors used to capture point clouds often have an inherent topology. Mobile Laser Scanners 
%(MLS) 
sample a regular grid in ($\theta, t$) where $\theta$ is the rotation angle of the laser beam and $t$ the instant of acquisition. Because the vehicle moves at a varying speed (to adapt to the traffic and respect the circulation rules) and may rotate,  the sampling is however not uniform in space.
%Because the number $N_p$ of pulses covering a 2$\pi$ rotation of the Laser beam is in general not integer, the regular sampling in $\theta$ yields a natural 6-neighborhood in $(\theta, t)$ that we call sensor topology 
In general, the number $N_{p}$ of pulses for a 2$\pi$ rotation in $\theta$ is not an integer so a pulse $P_i$ has six neighbors $P_{i-1}$, $P_{i+1}$, $P_{i-n}$, $P_{i-n-1}$, $P_{i+n}$, $P_{i+n+1}$ where $n=\lfloor N_p \rfloor$ is the integer part of {the number of pulses per line}
%$N_p$ 
as illustrated on figure \ref{fig:6v}.

However, this topology concerns emitted pulses, not recorded echoes. One pulse might have 0 echo (no target hit) or up to 8 as most modern scanners can record multiple echoes for one pulse if the laser beam intersected several targets, which is very frequent in the vegetation or transparent objects for instance.
% BV: Redondant For each pulse $P_i$, we assume that we know $N$ topological neighboring pulses $T(P) = \left\{P_{i_1}, P_{i_2}, ..., P_{i_N} \right\}$. $N$ can be 4 or 8 for regular grids, 6 for 1D sensors with non integer number of pulses per line, where the neighbors of a pulse $P$ in the adjacent lines will be the pulse directly before and after the $\theta$ of $P$, and decrease towards borders of the scan.
We chose to tackle this issue by connecting an echo to each echoes of its pulses' neighbors as illustrated in Figure \ref{fig:multi-echo} because we should keep all possible edge hypotheses before filtering them. 

\subsection{Edge filtering}

For each pair of connected echoes in sensor topology (as defined above), we need a criteria to decide whether we should keep it in the reconstructed simplicial complex. We propose the following:

\begin{itemize}
\item $C_0$ regularity: we want to prevent forming edges between echoes when their euclidean distance is too high.
\item $C_1$ regularity: we want to favor edges when two collinear edges share an echo.
\end{itemize}

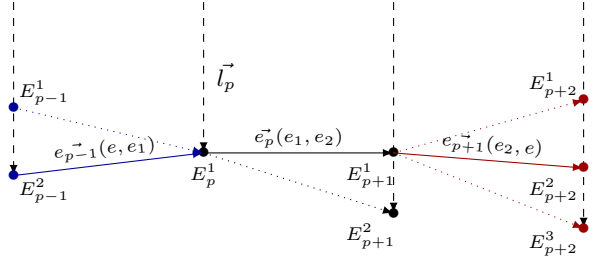
\begin{figure}[t]
\centering
\begin{tikzpicture}

%%% laser beams
\draw[-latex,black,thin,dashed] (0,2) -- (0,0);
\draw[-latex,black,thin,dashed] (2.5,2) -- (2.5,-.8);
\draw[-latex,black,thin,dashed] (5,2) -- (5,-1);
\draw[-latex,black,thin,dashed] (-2.5,2) -- (-2.5,-.3);

%%% echoes
\node at (0,0) {$\bullet$};
\node at (2.5,0) {$\bullet$};
\node at (2.5,-.8) {$\bullet$};
\node[red!60!black] at (5,-.2) {$\bullet$};
\node[red!60!black] at (5,.7) {$\bullet$};
\node[red!60!black] at (5,-1) {$\bullet$};
\node[blue!60!black] at (-2.5,-.3) {$\bullet$};
\node[blue!60!black] at (-2.5,.6) {$\bullet$};

%%% angle names
%\node at (0,-.4) {\scriptsize{$\beta_1$}};
%\node at (2.2,-.45) {\scriptsize{$\beta_2$}};

%%% echoes names
\node at (0,-.3) {\scriptsize{$E_p^{1}$}};
\node at (2.2,-.3) {\scriptsize{$E_{p+1}^{1}$}};
\node at (2.2,-1.1) {\scriptsize{$E_{p+1}^{2}$}};
\node at (-2.1,-.5) {\scriptsize{$E_{p-1}^{2}$}};
\node at (-2.1,.8) {\scriptsize{$E_{p-1}^{1}$}};
\node at (4.6,.9) {\scriptsize{$E_{p+2}^{1}$}};
\node at (4.6,-.5) {\scriptsize{$E_{p+2}^{2}$}};
\node at (4.6,-1.2) {\scriptsize{$E_{p+2}^{3}$}};

%%% adjacency
\draw[-latex,black] (0,0) -- (2.5,0);
\draw[-latex,black,dotted,thin] (0,0) -- (2.5,-.8);
\draw[-latex,red!60!black] (2.5,0) -- (5,-.2);
\draw[-latex,red!60!black,dotted,thin] (2.5,0) -- (5,-1);
\draw[-latex,red!60!black,dotted,thin] (2.5,0) -- (5,.7);
\draw[-latex,blue!60!black] (-2.5,-.3) -- (0,0);
\draw[-latex,blue!60!black,dotted,thin] (-2.5,.6) -- (0,0);

%%% angles
%\draw (2.2,0) arc (0:-180:.2);
%\draw (.2,0) arc (0:-175:.2);

%%% vectors
\node at (1.25,.2) {\scriptsize{$\vec{e_p}(e_1,e_2)$}};
\node[rotate=5] at (-1.3,.05) {\scriptsize{$\vec{e_{p-1}}(e,e_1)$}};
\node[rotate=-5] at (3.8,.1) {\scriptsize{$\vec{e_{p+1}}(e_2,e)$}};
\node at (.3,1) {$\vec{l_p}$};

\end{tikzpicture}
\caption{Illustration of the computed regularities $C_0$ and $C_1$. The black dots represent the echoes associated to the considered pulses. The blue and red ones correspond respectively to the precedent and following adjacent echoes. The solid arrows show the adjacent echoes used for the $C_0$ and $C_1$ computation. The black one correspond to the most orthogonal liaison to the sensor beams. The blue and red ones are selected because the angle between these vectors and the black one are the closest possible to $\pi$. The black dashed lines represent the laser beams.}
% * <bruno.vallet@ign.fr> 2018-01-05T16:30:13.937Z:
% 
% > Illustration of the computed regularities
% 1) tes rayons laser sont pas droits /ok
% 2) ca serait bien d'avoir 2 échos sur au moins un des deux pulses du centre /ok
% 3) Vires \beta_1 et \beta_2 /ok
% 4) Si possible ajoutes les E_p^e sur chaque echo /fait, pour moi ça reste encore lisible même si un pu chargé
% 
% ^.
\label{fig:regularities}
\end{figure}

In order to be independent from the sampling density, we propose to express the regularities in an angular manner. Moreover, the sensor topology has an hexagonal structure, and we propose to treat each line in the 3 directions of the structure independently. For the reminder of this article, and because a single pulse can have multiple echoes, we will express, for a pulse ${p}$, its echoes as $E_p^{e}$ where ${e} \in 1\ldots N_p$, with ${N_p}$ the number of echoes of ${p}$. We then express the regularities as:

\begin{itemize} 
% J'ai changé les e(p,e_1,e_2) par des e_p(e_1,e_2)
\item $C_0$ regularity, for an edge  $(E_{p}^{e_1}, E_{p+1}^{e_2})$ between two echoes of  two neighboring pulses: $$C_0(\mathit{p}, e_1, e_2)= 1 - \vec{e_p}(e_1, e_2)\cdot\vec{l_p} \quad,$$ where $ \vec{e_p}(e_1, e_2) = \frac{\overrightarrow{E_{p}^{e_1}E_{p+1}^{e_2}}}{||\overrightarrow{E_{p}^{e_1}E_{p+1}^{e_2}}||}$ and $\vec{l_p}$ is the direction of the laser beam of pulse $p$ (cf Figure \ref{fig:regularities}). $C_0$ is close to 0 for surfaces orthogonal to the LiDAR ray and close to 1 for grazing surfaces, almost parallel to the ray.
\item $C_1$ regularity, for an edge  $(E_{p}^{e_1}, E_{p+1}^{e_2})$ between two echoes of  two neighboring pulses:
\begin{equation}
\nonumber
\begin{split}
C_1(p, e_1, e_2) \!=\,& min_{e=1}^{N_{p-1}} |1-\vec{e}_{p-1}(e, e_1)\cdot \vec{e}_p(e_1, e_2)|\cdot \\ 
& min_{e=1}^{N_{p+2}} |1-\vec{e}_p(e_1, e_2)\cdot \vec{e}_{p+1}(e_2, e)|. 
\end{split}
\end{equation}
where the minima are given a value of 1 if the pulse is empty. $C_1$ is close to 0 is the edge is aligned with at least one of its neighboring edges, and close to 1 if it is orthogonal to all neighboring edges.
\end{itemize}

From these regularities, we propose a simple filtering based on the computed angles. 
%%% texte réécrit
Figure \ref{fig:regularities} illustrates the computation of the $C_0$ and $C_1$ regularities. Considering two adjacent echoes $E^{e_1}_p$ and $E^{e_2}_{p+1}$, the $C_0$ regularity can be interpreted as the cosine of the angle between the laser beam direction in $p$ and $\vec{e_p}(e_1,e_2)$. We want to favor low values of the $C_0$ as it corresponds to echoes with a low depth difference. On the other side, to compute the $C_1$ regularity, we have to browse the echoes of the preceding and following pulses along the 3 directions of our structure. For the preceding and following pulses, we select the echo which minimizes $\mid 1 - \vec{e_{p-1}}(e,e_1) \cdot \vec{e_p}(e_1,e_2) \mid$ (respectively $\mid 1 - \vec{e_{p+1}}(e_2,e) \cdot \vec{e_p}(e_1,e_2) \mid$). This gives us an information about the tendency of the considered edge to be collinear with at least one of its adjacent edges. We will favor the most collinear cases.

%For the reminder of this article, we denote $\alpha$ the angle corresponding to: $(\widehat{\vec{l_p},\vec{E_{p}^{e_1},E_{p+1}^{e_2}}})$ and $\beta$ the angle corresponding to: $(\widehat{\vec{E_{p}^{e_1},E_{p+1}^{e_2}},\vec{E_{p+1}^{e_2},E_{p+2}^{e}}})$ (or similarly: $(\widehat{\vec{E_{p-1}^{e},E_{p}^{e_1}},\vec{E_{p}^{e_1},E_{p+1}^{e_2}}})$). We want to favor the cases where $\alpha \simeq \pi/2$ (i.e. $C_0 \simeq 1$) because it represents the case where the echoes are orthogonal to the laser beam. On the other side, we want to encourage values of $\beta$ close to $\pi$ (i.e. $C_1 \simeq 0$) as it corresponds to a triplet of points aligned.

\begin{figure}[t]
\centering
\begin{tikzpicture}[scale=0.6]

\tikzset{
        hatch distance/.store in=\hatchdistance,
        hatch distance=10pt,
        hatch thickness/.store in=\hatchthickness,
        hatch thickness=.5pt
    }

    \makeatletter
    \pgfdeclarepatternformonly[\hatchdistance,\hatchthickness]{flexible hatch}
    {\pgfqpoint{0pt}{0pt}}
    {\pgfqpoint{\hatchdistance}{\hatchdistance}}
    {\pgfpoint{\hatchdistance-1.5pt}{\hatchdistance-1.5pt}}%
    {
        %\pgfsetcolor{\tikz@pattern@color}
        \pgfsetlinewidth{\hatchthickness}
        \pgfpathmoveto{\pgfqpoint{0pt}{0pt}}
        \pgfpathlineto{\pgfqpoint{\hatchdistance}{\hatchdistance}}
        \pgfusepath{stroke}
    }
    \makeatother

  \begin{axis}[ 
  	ytick style={draw=none},
    xtick style={draw=none},
  	xmin=0, xmax=1, ymin=0, ymax=1,
    samples=100,
    xticklabels=\empty,%{},
    extra x ticks={0,1},
	extra x tick labels={0,1},
    yticklabels={},
    extra y ticks={0,1},
	extra y tick labels={0,1},
    xlabel=\large{$C_0$},
    ylabel=\begin{turn}{270}\large{$C_1$}\end{turn}
  ]

    \addplot+[mark=none,
    	draw=none,
    	color=black!60,
        domain=0.59:1,
        samples=100,
        pattern=flexible hatch]{-.22+0.1*-x/(0.55-x)} \closedcycle;
    \addplot+[mark=none,
    	draw=none,
    	color=black!60,
        domain=0:0.59,
       	samples=100,
        pattern=flexible hatch]{11} \closedcycle;
        
    \addplot [mark=none,domain=0.59:1,color=blue]{-.22+0.1*-x/(0.55-x)}; 
    \draw [red, thin] (59,0) -- (59,115);
  \end{axis}
  
  \node at (4.05,-.1) {\scriptsize{$\mid$}};
  \node at (4.05,-.4) {\scriptsize{$\alpha_m$}};
\end{tikzpicture}
\caption{Filtering of edges knowing $\protect C_0$ and $\protect C_1$. The red line corresponds to the $\alpha_m$ threshold on $C_0$. The hatched area corresponds to the edges that we keep.}
% * <bruno.vallet@ign.fr> 2018-01-05T16:36:50.464Z:
% 
% > Filtering of edges knowing $\protect C_0$ and $\protect C_1$. The red line corresponds to the threshold on $C_0$. The hatched area corresponds to the edges that we keep.
% 1) Tu peux laisser \alpha_m au lieu de C_0 threshold. /ok
% 2) Comme on a dit que C_0=0 dans le cas régulier (orthogonal) il faut faire un mirroir horizontal de ta figure. C'est bcp plus logique comme ça, on verra qu'on rejette les edges pour lesquelles C0 et C1 sont grands. /ok
% 3) C0 et C1 sont entre 0 et 1 donc met des 0 et des 1 sur les axes et rend ton rectangle carré. /ok
% 
% 
% ^.
\label{fig:thresh}
\end{figure}
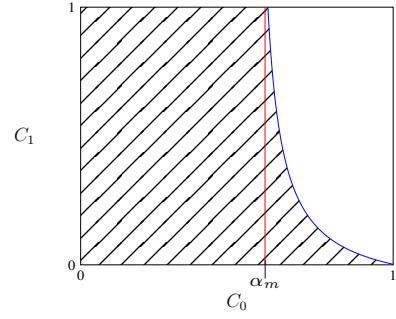

%Given two echoes, we consider that if $\alpha$ is high enough, the echoes belong to the same object, thus we don't have to compute $\beta$ with preceding or following echoes. We denote the threshold above which we assume echoes belong to same object: $\alpha_m$. However, if $\alpha$ is not high enough, we compute $\beta$ and filter the edges according to the value of $\alpha$ and $\beta$ when:
Given two adjacent echoes, we consider that if the $C_0$ regularity is high enough, we can ensure the real existence of the edge, and don't have to compute the $C_1$ regularity. We denote this threshold $\alpha_m$. For all the other cases, we compute the $C_1$ regularity and filter the edges according to the $C_0$ and $\alpha_m$:
$$
C_1 < \frac{\lambda \cdot \alpha_m \cdot C_0}{\alpha_m - C_0}\quad,
$$
where $\lambda$ sets how much $C_1$ regularity can compensate for $C_0$ discontinuity. A high value of $\lambda$ allows more edges to be kept.
This criteria is illustrated on figure \ref{fig:thresh}. The red line represent the $\alpha_m$ threshold and the blue line corresponds to the limit cases between removing and keeping the edges depending on $C_0$ and $C_1$.

We also want to filter edges that would remain single in the cloud, or just connected to one other edge but with different directions. Actually this often occurs on noisy areas where an edge can pass the regularity criteria "by chance", but it is very unprobable that this happens for two neighboring edges. That's why we propose an additional criterion to favor a reconstruction that leaves points instead of isolated or unaligned edges.

% e minuscules pour cohérence de notation
Let $e$ be an edge and $\left\{e_1, \ldots, e_n\right\}$ its adjacent edges. We consider that if we find an edge $e_i \in \left\{e_1, \ldots, e_n\right\}$ so that:
%We express this filtering similarly to the triangle filtering, but on all remaining edges. This time, we browse all adjacent edges $\left\{E_1, \ldots, E_n\right\}$ to a single edge $E$, and if we find an edge $E_i \in \left\{E_1, \ldots, E_n\right\}$ so that: 

$$
1 - \vec{e} \cdot \vec{e_i} < \epsilon\quad,
$$
where $\epsilon$ is the tolerance on edge reconstruction, $e$ is not alone or unaligned and we keep it in the simplicial complex.

\subsection{Triangle filtering}

Once we obtained a set of edges in our point cloud, a simple approach to filter triangles is to keep only the triangles (from sensor topology) which three edges have survived the edge filtering described above. Even if this method is an easy way to retrieve most triangles of the scene, it prevents recovering triangles in areas where edges are close to the threshold, in which case triangles will often have some edges just below and some just above the threshold so most triangles will be filtered out. 

In order to regularize the triangulation computed, we want to favor triangles that are coplanar to some of their adjacent triangles, in the same way we favored edges aligned with at least one neighboring edge. This is motivated by the fact that we want to ensure spatial regularity in our scene. Moreover, we found very unlikely the cases where a triangle is left alone. Also we want to remove all the triangles that may be formed by edges in noisy parts of the cloud as we cannot ensure their existence in the real scene. 

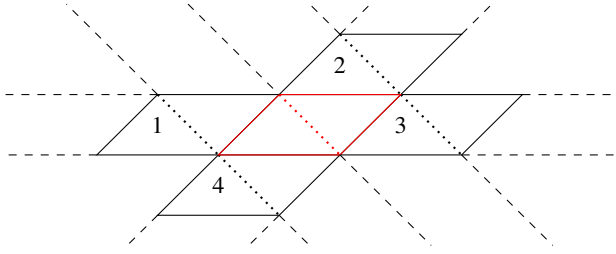
\begin{figure}[t]
\centering
\begin{tikzpicture}[scale=.8]
\draw [thin] (0,0) -- (1,1) -- (-1,1) -- (-2,0) -- cycle;
\draw [thick,dotted] (-1,1) -- (0,0);

\draw [thin] (4,0) -- (5,1) -- (3,1) -- (2,0) -- cycle;
\draw [thick,dotted] (3,1) -- (4,0);

\draw [thin] (2,2) -- (1,1) -- (3,1) -- (4,2) -- cycle;
\draw [thick,dotted] (2,2) -- (3,1);

\draw [thin] (0,0) -- (-1,-1) -- (1,-1) -- (2,0) -- cycle;
\draw [thick,dotted] (0,0) -- (1,-1);

\draw [thin,red] (0,0) -- (1,1) -- (3,1) -- (2,0) -- cycle;
\draw [thick,dotted,red] (1,1) -- (2,0);

\draw [thin,dashed] (-2,0) -- (-3.5,0);
\draw [thin,dashed] (-1,1) -- (-3.5,1);
\draw [thin,dashed] (5,1) -- (6.5,1);
\draw [thin,dashed] (4,0) -- (6.5,0);
\draw [thin,dashed] (2,2) -- (2.5,2.5);
\draw [thin,dashed] (4,2) -- (4.5,2.5);
\draw [thin,dashed] (-1,-1) -- (-1.5,-1.5);
\draw [thin,dashed] (1,-1) -- (0.5,-1.5);
\draw [thin,dashed] (-.5,2.5) -- (1,1);
\draw [thin,dashed] (2,0) -- (3.5,-1.5);
\draw [thin,dashed] (-1,1) -- (-2.5,2.5);
\draw [thin,dashed] (1,-1) -- (1.5,-1.5);
\draw [thin,dashed] (2,2) -- (1.5,2.5);
\draw [thin,dashed] (4,0) -- (5.5,-1.5);

\node at (-1,0.5) {1};
\node at (2,1.5) {2};
\node at (3,0.5) {3};
\node at (0,-0.5) {4};
\end{tikzpicture}
\caption{Representation of a wedge (red) and its adjacent wedges. The limit between the triangles of each wedge is represented with a dotted line. The dashed lines stand for the directions of our structure. Wedges adjacent to the red one are numbered from 1 to 4.}
\label{fig:wedge_adj}
\end{figure}

As triangles are 2D objects, we want to define a 2D $C_1$ regularity by separating between $C_1$ regularity along two directions. Unfortunately, a triangle has 3 neighbors. We solve the problem by filtering pairs of adjacent triangles (that we will call wedges) which have four adjacent wedges in 2 separate directions as illustrated on figure \ref{fig:wedge_adj}.

The filtering we propose to keep the wedges that are $C_1$ regular with neighboring wedges in the two directions, where $C_1$ regularity is defined as the criterion:
$$
1-|W^N \cdot W_i^N| < \omega\quad,
$$
where $W$ is a wedge whose normal is $W^N$ and $\left\{W_1, \ldots, W_n\right\}$ are its adjacent wedges whose normals are $\left\{W_1^N, \ldots, W_n^N\right\}$ respectively. This means, on figure \ref{fig:wedge_adj}, that if the red wedge is only $C_1$ regular with the wedges 1 and 3, it will be discarded, while it will be kept if it is regular with only 1 and 2. $\omega$ is the tolerance on the co-planarity of the two wedges to define regularity. The rationale behind this choice is the same as for the edges: being irregular with both neighbors in one direction means that we are on a depth discontinuity in that direction that cannot be distinguished from a grazing surface, while regularity with at least one neighbor in both directions means that the wedge is part of a (potentially grazing) planar surface.

\section{Results}

We implemented the pipeline presented before, first with only the edge filtering and the simple triangle reconstruction from edge effectively forming a triangle. Then we added the triangle filtering part. We compared our results with a naive filtering on edge length where triangles in the simplicial complex correspond to all triplets of edges forming a triangle.

For all the following tests, we used data from the Stereopolis vehicle \citep{paparoditis2012stereopolis}. The scenes have been acquired in an urban environment (Paris) {and are mostly composed of roads, facades, trees and urban planning}. All the simplicial complexes presented in this part will be represented as follow:
\begin{itemize}
\item triangles in red,
\item edges that are not part of any triangle in green,
\item points that don't belong to any triangle or edge in black.
\end{itemize}
Note that following its mathematical definition, the endpoints of an edge of a simplicial complex also belong to the complex, and similarly for the edges of a triangle, but we do not display them for clarity.

The parameter search phase was conducted in two experiments. In the first set of experiments, the impact of parameters $\alpha_m$ and $\lambda$ were studied. For the remaining parameters $\omega$ and $\epsilon$, their influence was tested in a second batch of experiments. Because all our criteria depend on trigonometric functions (the dot products of normalized vectors is the cosine of their angle), all these parameters are to choose in $[0,1]$. Last, we compare both methods with the naive filtering on edge length.

\subsection{Parametrization of $\alpha_m$ and $\lambda$}

\begin{figure*}[t]
\centering
\begin{subfigure}[t]{0.2\textwidth}
\centering
\includegraphics[width=1\textwidth]{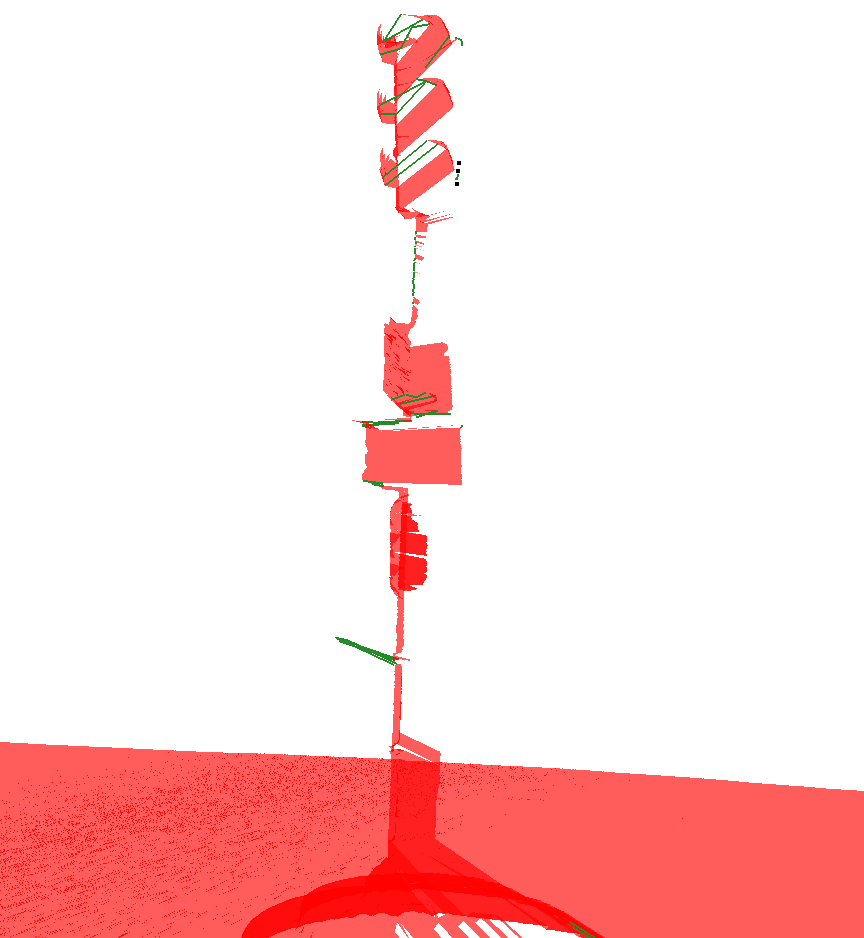}
\caption{$\alpha_m$= 0.01}
\end{subfigure}
\hspace{.1cm}
\begin{subfigure}[t]{0.2\textwidth}
\centering
\includegraphics[width=1\textwidth]{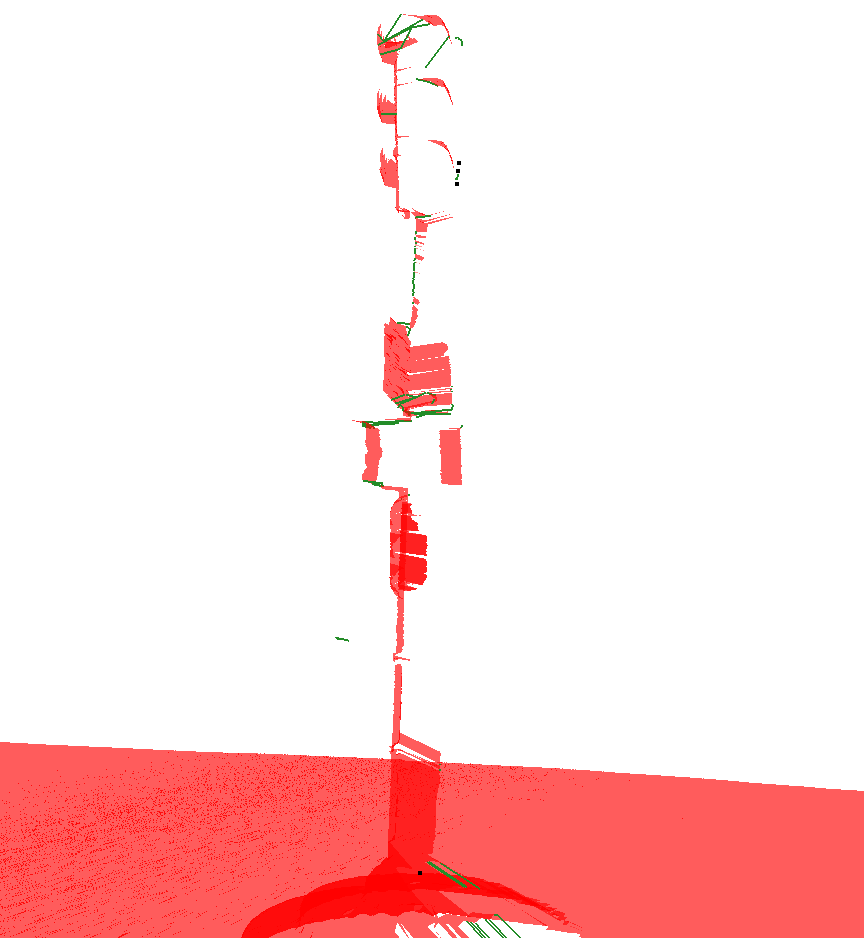}
\caption{$\alpha_m$= 0.05}
\end{subfigure}
\hspace{.1cm}
\begin{subfigure}[t]{0.2\textwidth}
\centering
\includegraphics[width=1\textwidth]{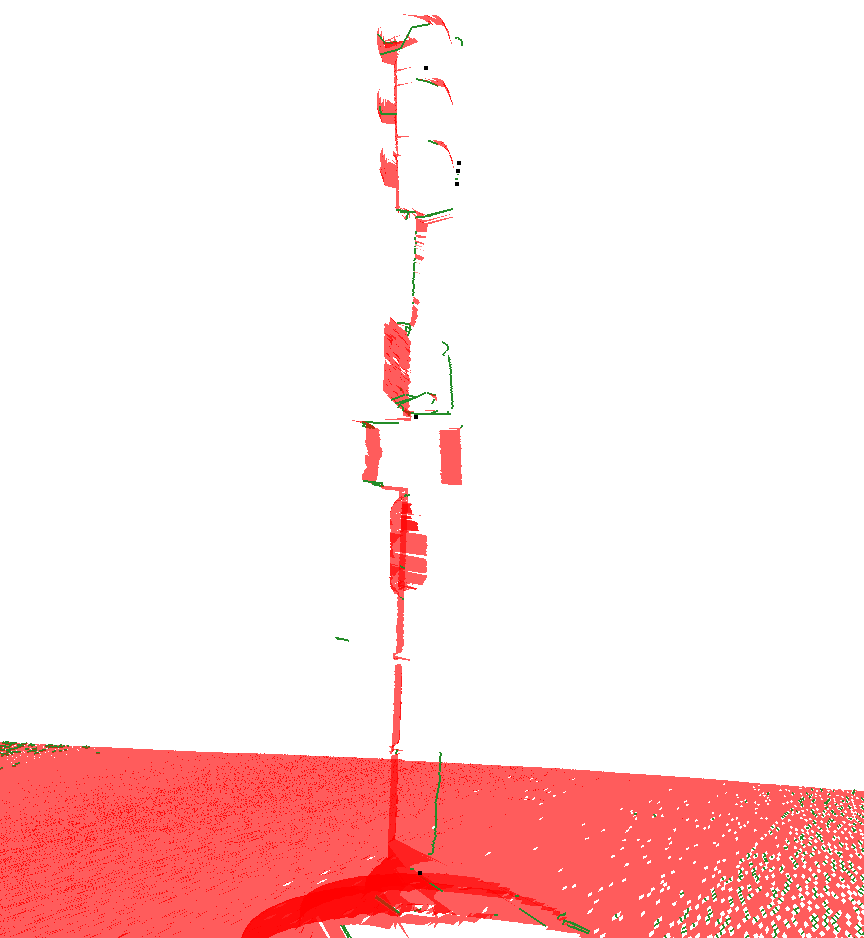}
\caption{$\alpha_m$= 0.1}
\end{subfigure}
\hspace{.1cm}
\begin{subfigure}[t]{0.2\textwidth}
\centering
\includegraphics[width=1\textwidth]{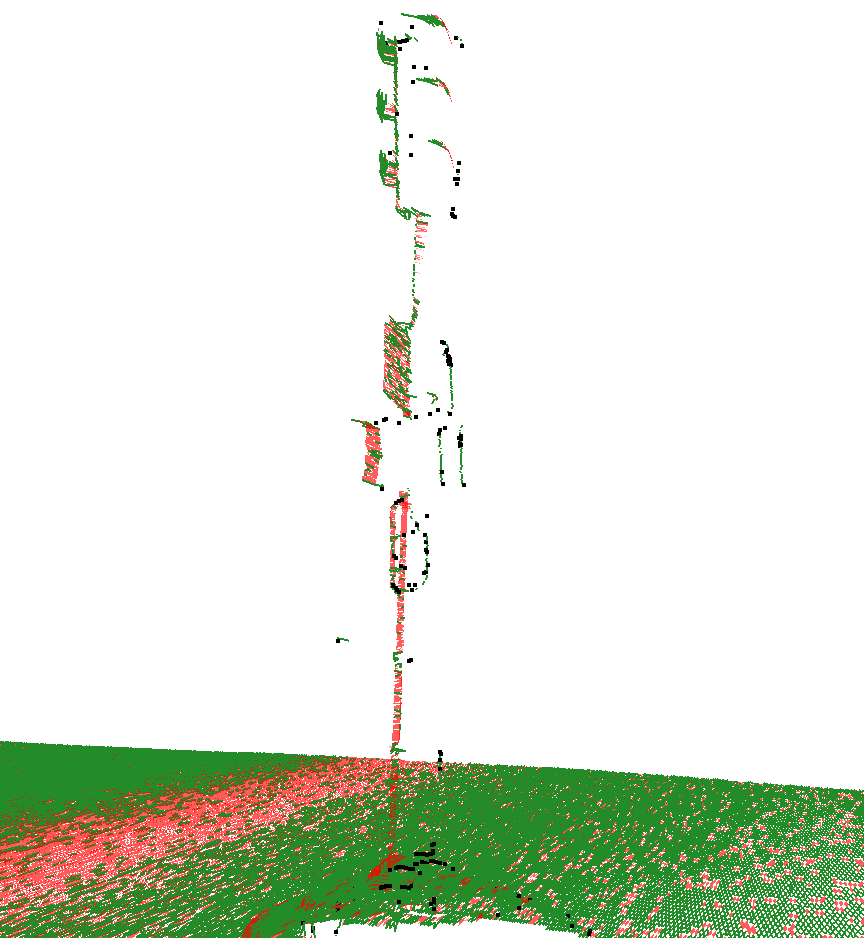}
\caption{$\alpha_m$= 0.5}
\end{subfigure}
\caption{Influence of $\alpha_m$. $\lambda$ is fixed to $10^{-4}$.}
\label{fig:alpha_m}
\end{figure*}

We first studied the influence of $\alpha_m$. A high value will discard a lot of edges and prevent the formation of triangles, whereas a low value will preserve too many edges on real discontinuities. The results are presented in figure \ref{fig:alpha_m}. We see on the left example that on the one hand, low values of $\alpha_m$ allow the formation of edges between the bottom of the traffic sign and the road. On the other hand, high values of $\alpha_m$ show that the reconstruction of triangles is harder, even on the road.

\begin{figure*}[t]
\centering
\begin{subfigure}[t]{0.2\textwidth}
\centering
\includegraphics[width=1\textwidth]{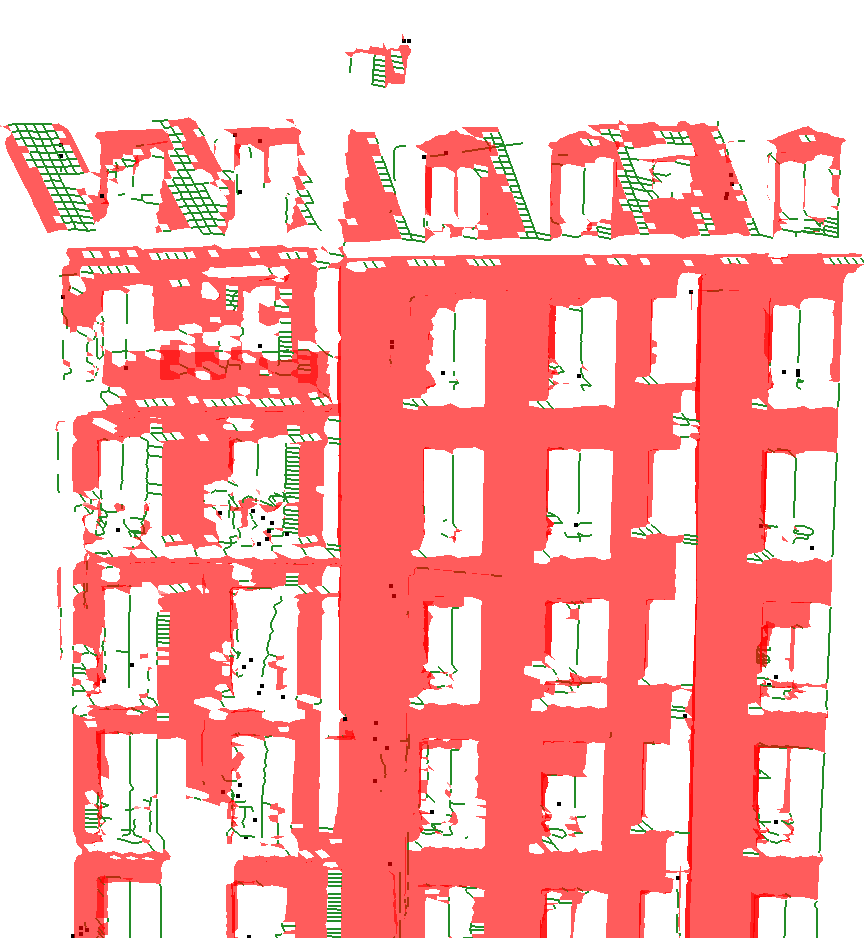}
\caption{$\lambda = 10^{-6}$}
\end{subfigure}
\hspace{.1cm}
\begin{subfigure}[t]{0.2\textwidth}
\centering
\includegraphics[width=1\textwidth]{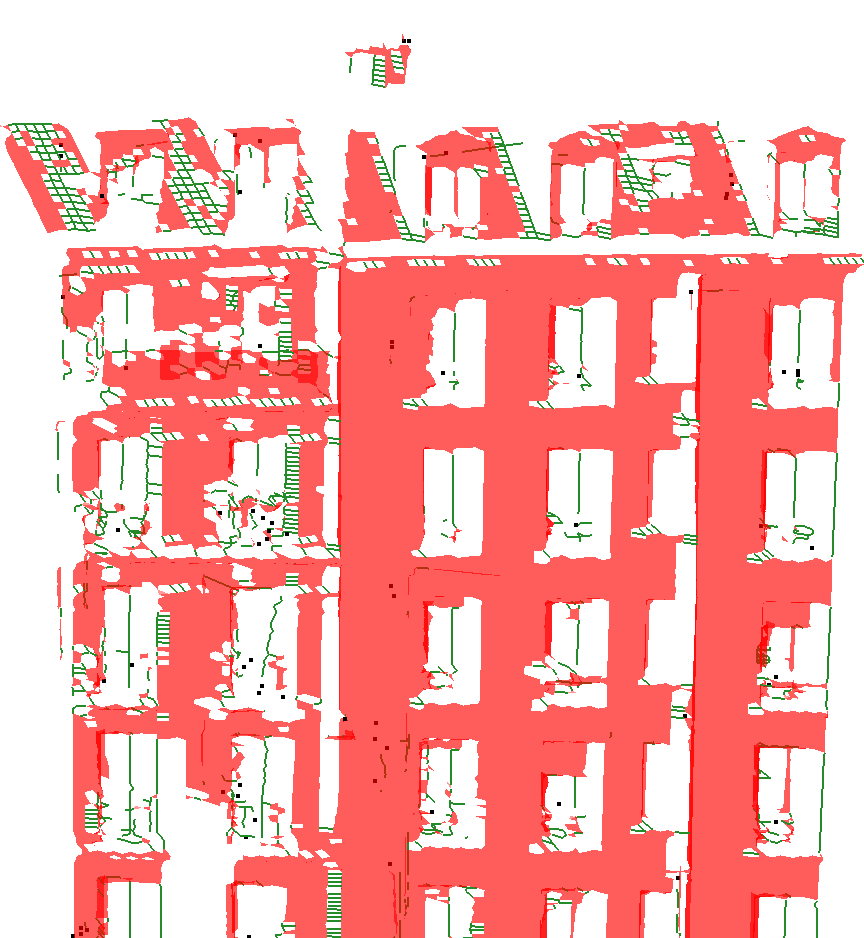}
\caption{$\lambda = 10^{-4}$}
\end{subfigure}
\hspace{.1cm}
\begin{subfigure}[t]{0.2\textwidth}
\centering
\includegraphics[width=1\textwidth]{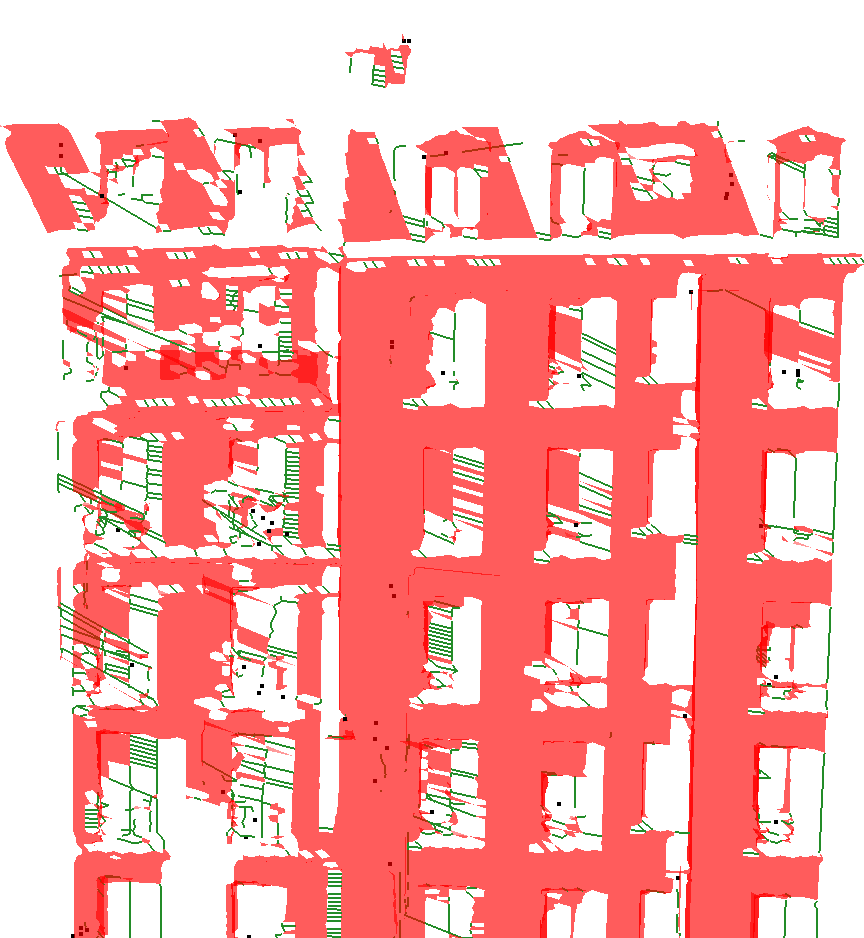}
\caption{$\lambda = 10^{-2}$}
\end{subfigure}
\hspace{.1cm}
\begin{subfigure}[t]{0.2\textwidth}
\centering
\includegraphics[width=1\textwidth]{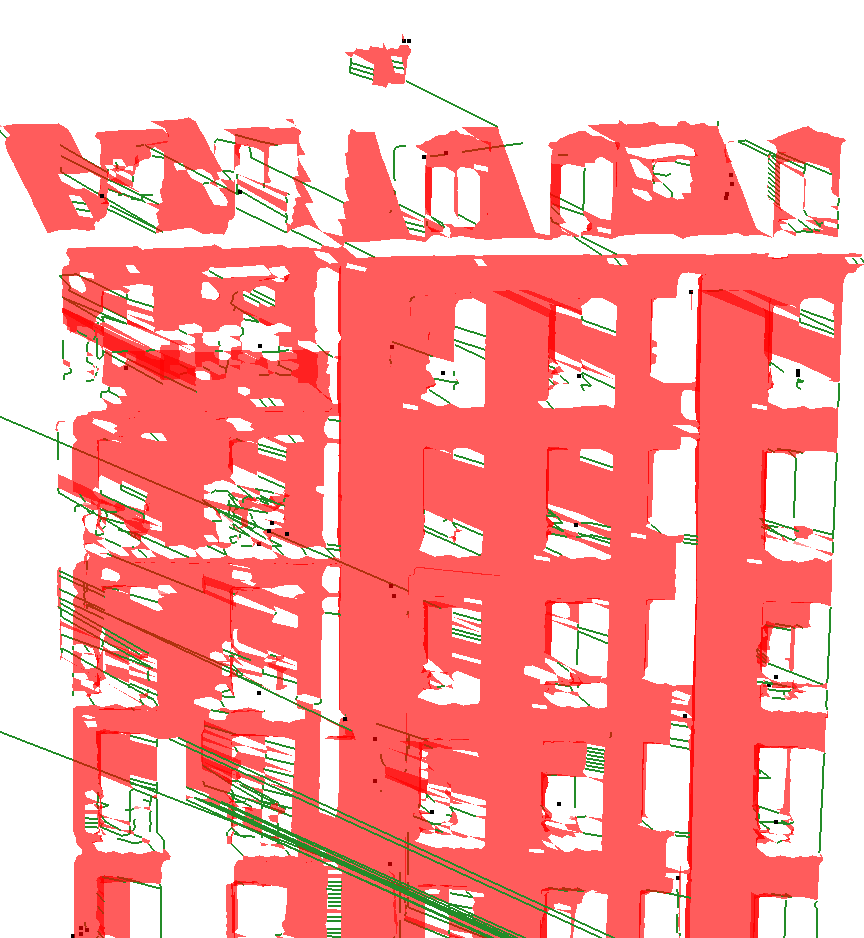}
\caption{$\lambda = 0.1$}
\end{subfigure}
\caption{Influence of $\lambda$. $\alpha_m$ is fixed to 0,05. The scene represents a facade in the grazing surface case.}
\label{fig:lambda}
\end{figure*}

The second parameter of this method, $\lambda$ corresponds to the fact that we want to preserve grazing surfaces where edges are long (important depth discontinuity) but collinear. Figure \ref{fig:lambda} illustrates the tuning of this $\lambda$ parameter. On the one hand, for high values of $\lambda$ (right), {edges between window bars and walls or insides of buildings are created.}
%edges between the person and the building are created. 
On the other hand, when $\lambda$ is too low (left), only the best edges are retrieved. 
%We note that there is nearly no difference between the two lowest values of $\lambda$ tested here, except for the part of the building occluded by people. 
For these cases, the number of remaining edges is low (hundreds of edges for millions of points), and lowering $\lambda$ removes edges that may be useful for human interpretation of the reconstruction.

% BV: TODO: j'aurais préféré voir une surface razante pour voir l'impact de \lambda, vu qu'on justifie ce critère comme ça

\subsection{Parametrization of $\omega$ and $\epsilon$}

\begin{figure*}[t]
\centering
\begin{subfigure}[t]{0.2\textwidth}
\centering
\includegraphics[width=1\textwidth]{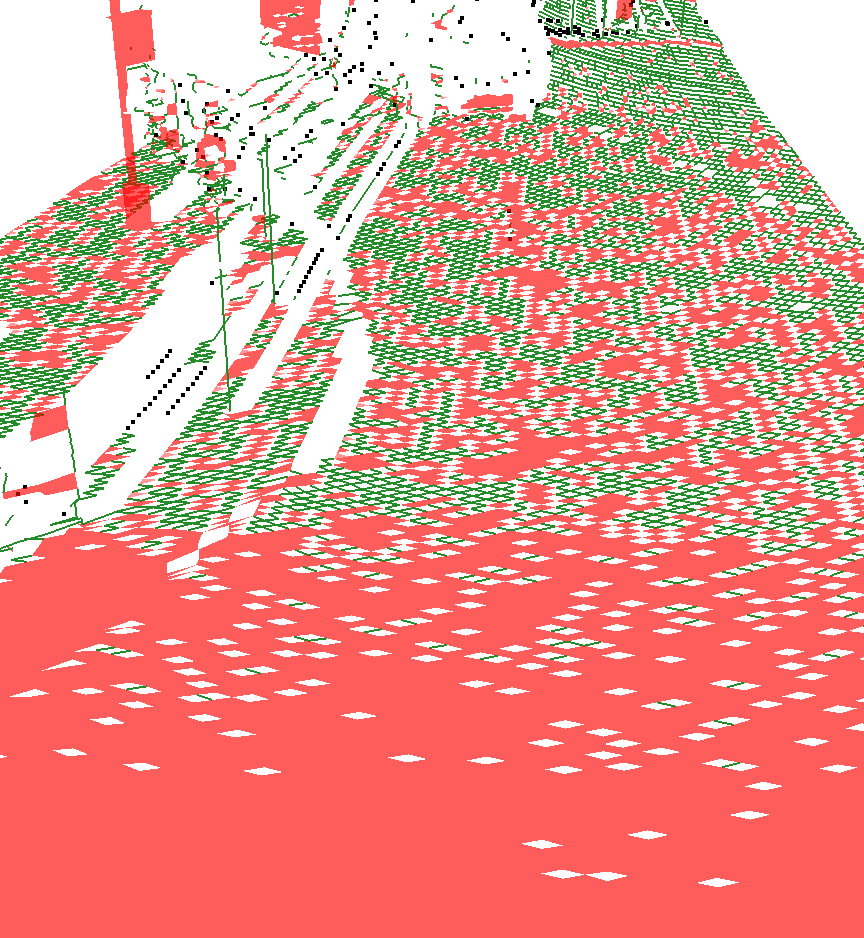}
\caption{No triangle filtering}
\end{subfigure}
\hspace{.1cm}
\begin{subfigure}[t]{0.2\textwidth}
\centering
\includegraphics[width=1\textwidth]{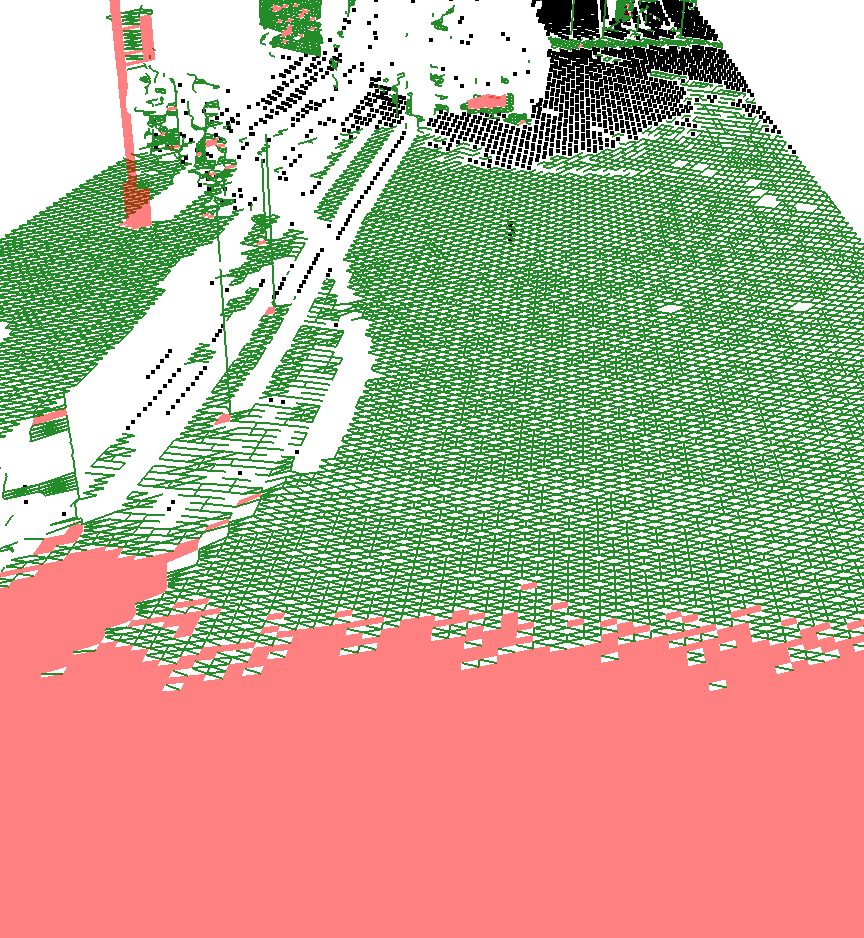}
\caption{$\omega = 10^{-5}$}
\end{subfigure}
\hspace{.1cm}
\begin{subfigure}[t]{0.2\textwidth}
\centering
\includegraphics[width=1\textwidth]{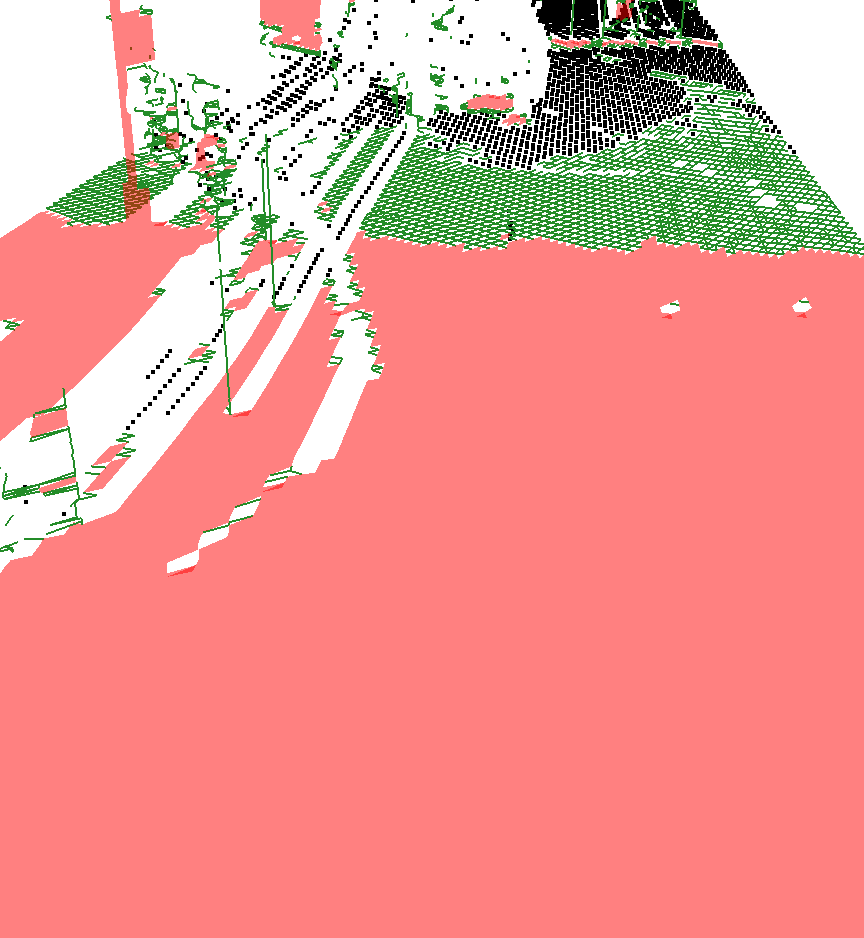}
\caption{$\omega = 10^{-4}$}
\end{subfigure}
\hspace{.1cm}
\begin{subfigure}[t]{0.2\textwidth}
\centering
\includegraphics[width=1\textwidth]{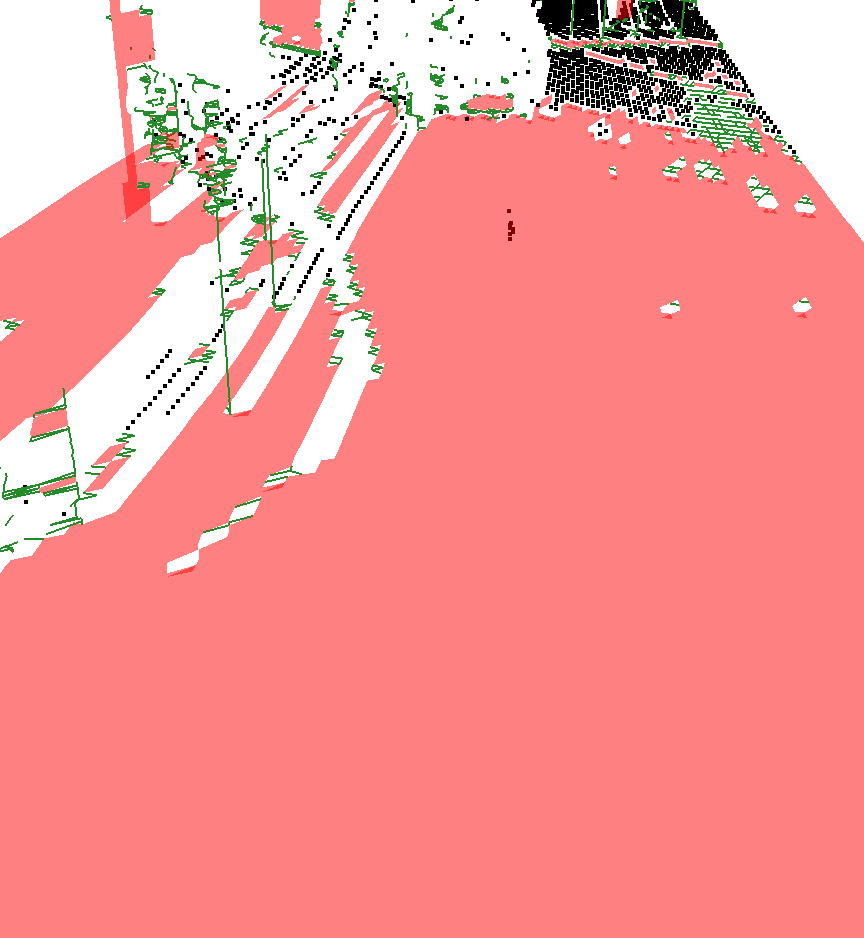}
\caption{$\omega = 10^{-3}$}
\end{subfigure}
\caption{Parametrization of $\omega$. $\epsilon$ is fixed to $5\cdot10^{-3}$. The scene represents a road in the grazing surface case.}
\label{fig:omega}
\end{figure*}

For this set of experiments, $\alpha_m$ and $\lambda$ were fixed respectively to $0.05$ and $10^{-4}$. The influence of the last two parameters is especially visible on noisy areas and grazing surfaces, where the level of detail of the scene is close to the acquisition density. We first studied the effect of parameter $\omega$ on triangles. For high values of $\omega$, we expect that a lot of triangles will be retrieved, especially in the grazing surface case, where sometimes our algorithm struggles to retrieve edges in the 3 directions (but performs well on two directions). The results are shown on figure \ref{fig:omega} and illustrate our problem in the grazing surface case. The figure on the left is the baseline computed previously. As expected, the number of triangles increase for high values of $\omega$ and the road is cleaner than without the triangle filtering part. Its main drawback is its propensity to let a few triangles in noisy areas like tree's foliage.
%For low values of $\omega$ the number of triangles decreases, therefore increasing the number of (visible) edges. The results are shown on figure \ref{fig:omega}. The figure on the left is the baseline computed previously. As expected, the triangles in the noisy part at the top and bottom of the doors are mainly replaced with edges. A side effect of this parameter is its tendency to add holes in surfaces that were well retrieved before. This is particularly visible on the figure on the right.

\begin{figure*}[t]
\centering
\begin{subfigure}[t]{0.2\textwidth}
\centering
\includegraphics[width=1\textwidth]{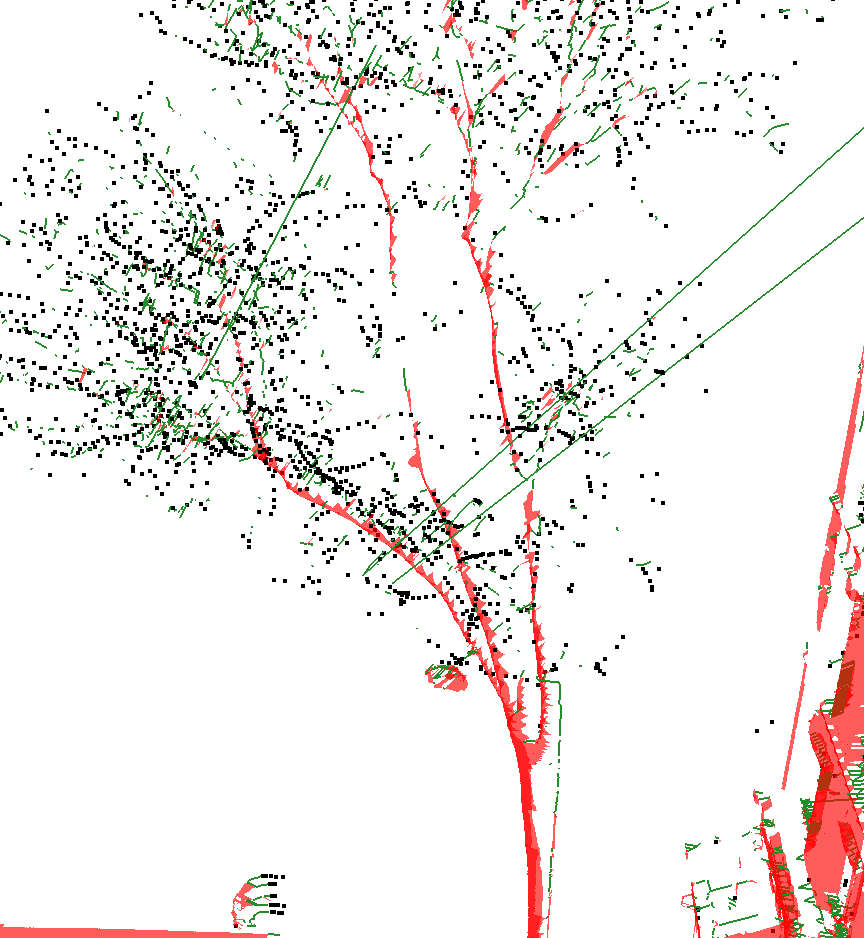}
\caption{No triangle filtering}
\end{subfigure}
\hspace{.1cm}
\begin{subfigure}[t]{0.2\textwidth}
\centering
\includegraphics[width=1\textwidth]{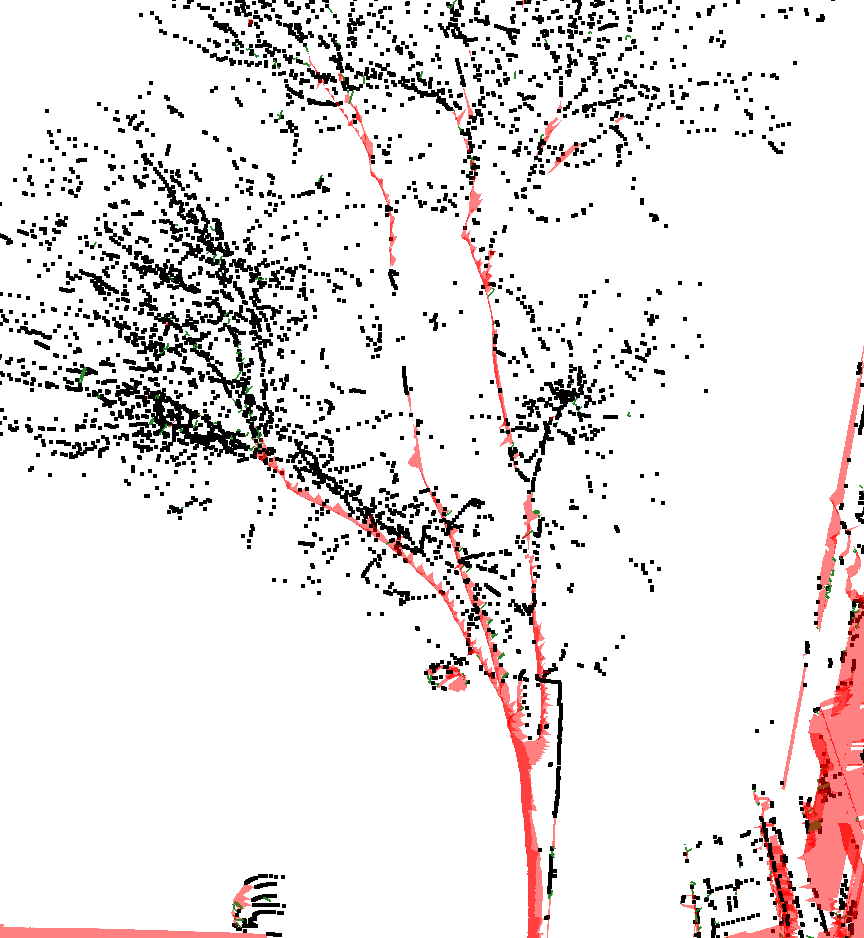}
\caption{$\epsilon = 10^{-5}$}
\label{fig:epsilon_10-5}
\end{subfigure}
\hspace{.1cm}
\begin{subfigure}[t]{0.2\textwidth}
\centering
\includegraphics[width=1\textwidth]{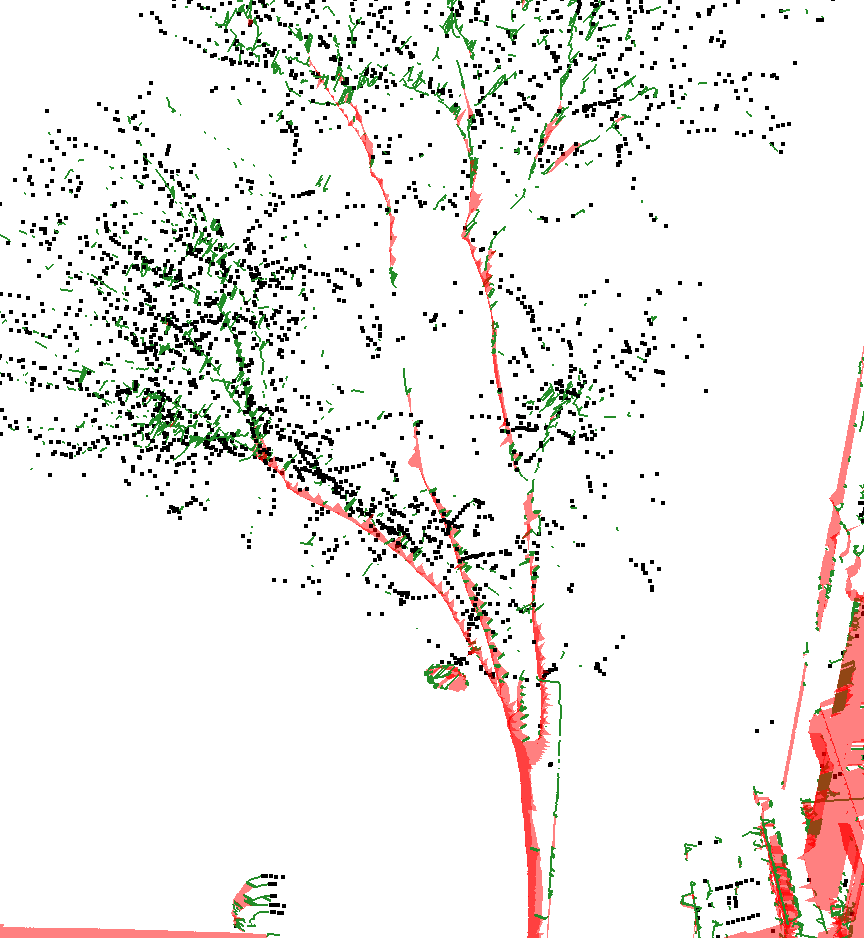}
\caption{$\epsilon\!\!=\!\!5\cdot10^{-3}$}
\end{subfigure}
\hspace{.1cm}
\begin{subfigure}[t]{0.2\textwidth}
\centering
\includegraphics[width=1\textwidth]{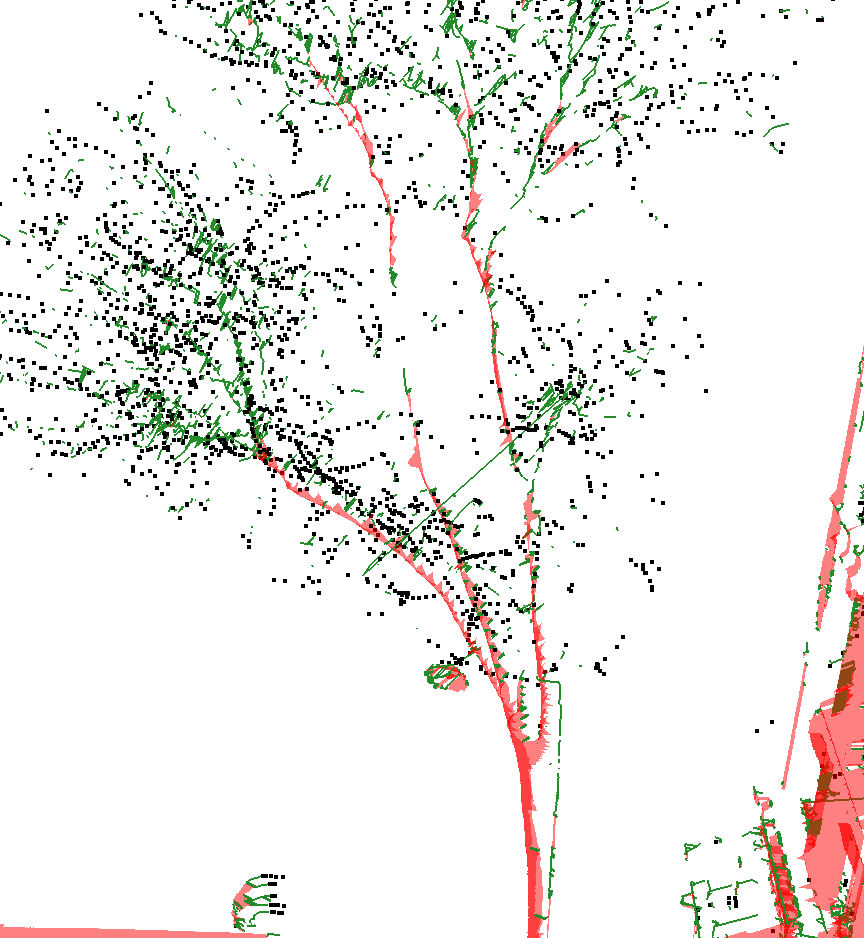}
\caption{$\epsilon = 10^{-2}$}
\label{fig:epsilon_10-2}
\end{subfigure}
\caption{Parametrization of $\epsilon$. $\omega$ is fixed to $10^{-3}$. The scene represents a tree with its foliage.}
\label{fig:epsilon}
\end{figure*}

The second parameter $\epsilon$ is a regularization term on the edges. Low values of $\epsilon$ will decrease the number of edges, thus leaving many points not linked to others. Results are presented in figure \ref{fig:epsilon}. As previously, the figure on the left shows the output of the first filtering step. Figures corresponding to lowest values of $\epsilon$ respect our previsions: the number of edges keeps decreasing whereas the number of points increase. A side effect of this method can be seen on figure \ref{fig:epsilon_10-5}, as there is nearly no edge in the tree's foliage.
%a line of points acquired on the extremity of the facade was rendered by a line of edges in \ref{fig:epsilon_10-2} whereas most edges have been discarded in \ref{fig:epsilon_10-3}, leaving only points.

\subsection{Comparison of the three methods}

In this part, we compare our two methods to a naive filtering based on edge lengths alone. The naive filtering is based on a $0.5$ meters threshold. For both methods, $\alpha_m$ and $\lambda$ are respectively fixed to $0.05$ and $10^{-4}$. Furthermore, for the last method, $\omega$ and $\epsilon$ are fixed to $10^{-3}$ and $5\cdot10^{-3}$ respectively. A video of the results is available at \citep{video}.
%\STF{The computational time for the naive reconstruction method is approximately 0.88 seconds, 5.6 seconds for the edge filtering and 75 seconds with the triangle filtering.}

\begin{figure*}[!h]
\centering
\begin{subfigure}[t]{0.435\textwidth}
\centering
\includegraphics[width=0.95\textwidth,height=0.81\textwidth]{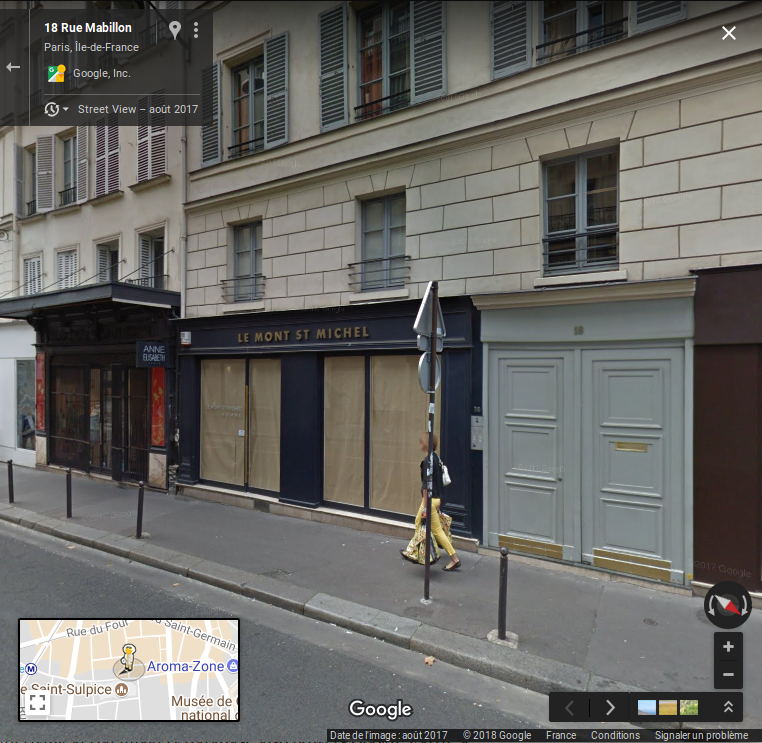}
%%% redimensionnement pour ajuster la taille (les autres images font 964 * 754 pixels)
\caption{Image of the scene from \citet{streetview}}
\end{subfigure}
\begin{subfigure}[t]{0.435\textwidth}
\centering
\includegraphics[width=0.95\textwidth,height=0.81\textwidth]{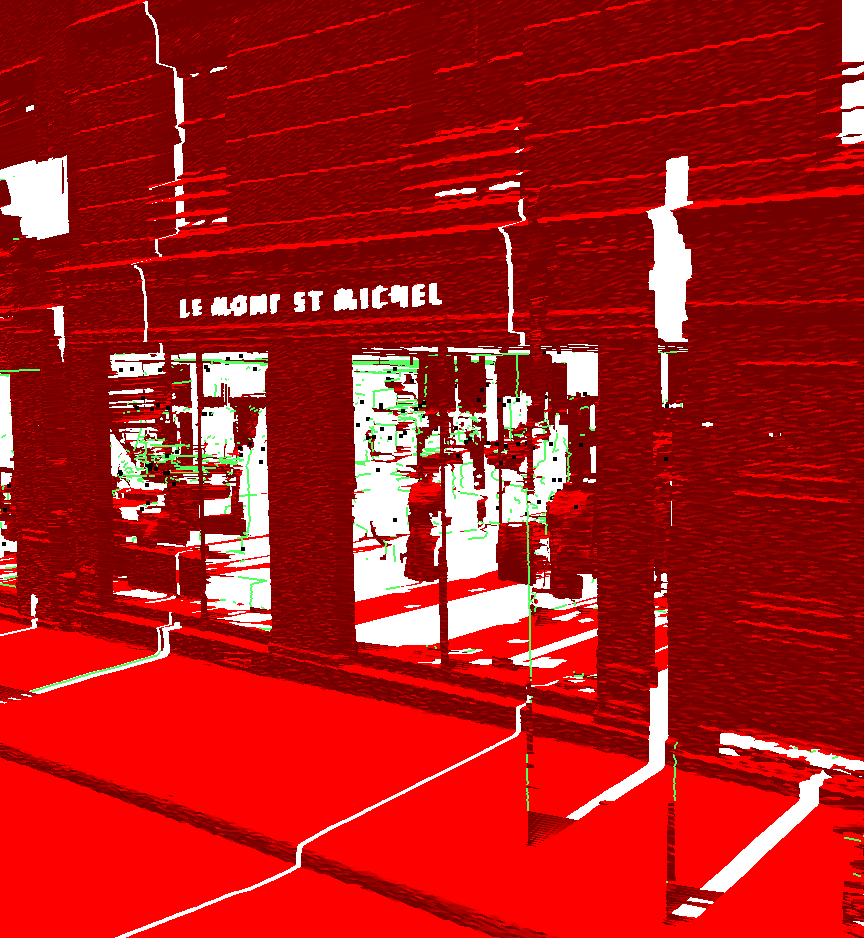}
\caption{Naive method}
\end{subfigure}
\begin{subfigure}[t]{0.435\textwidth}
\centering 
\includegraphics[width=0.95\textwidth,height=0.81\textwidth]{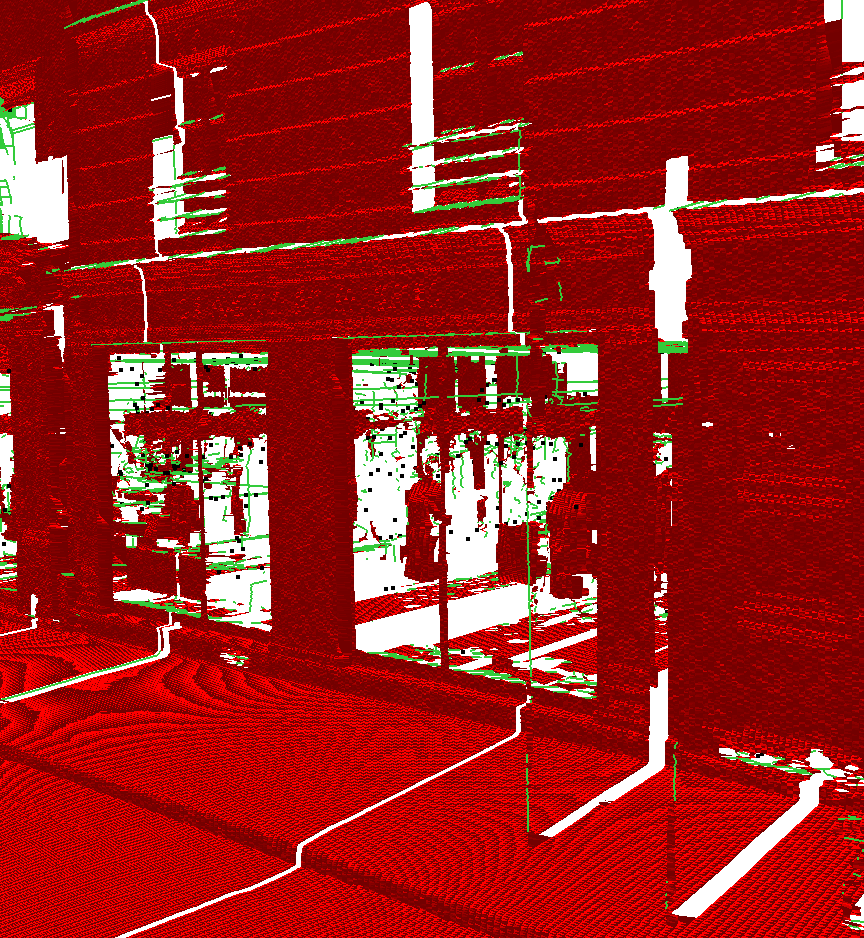}
\caption{Edge filtering}
\end{subfigure}
\begin{subfigure}[t]{0.435\textwidth}
\centering
\includegraphics[width=0.95\textwidth,height=0.81\textwidth]{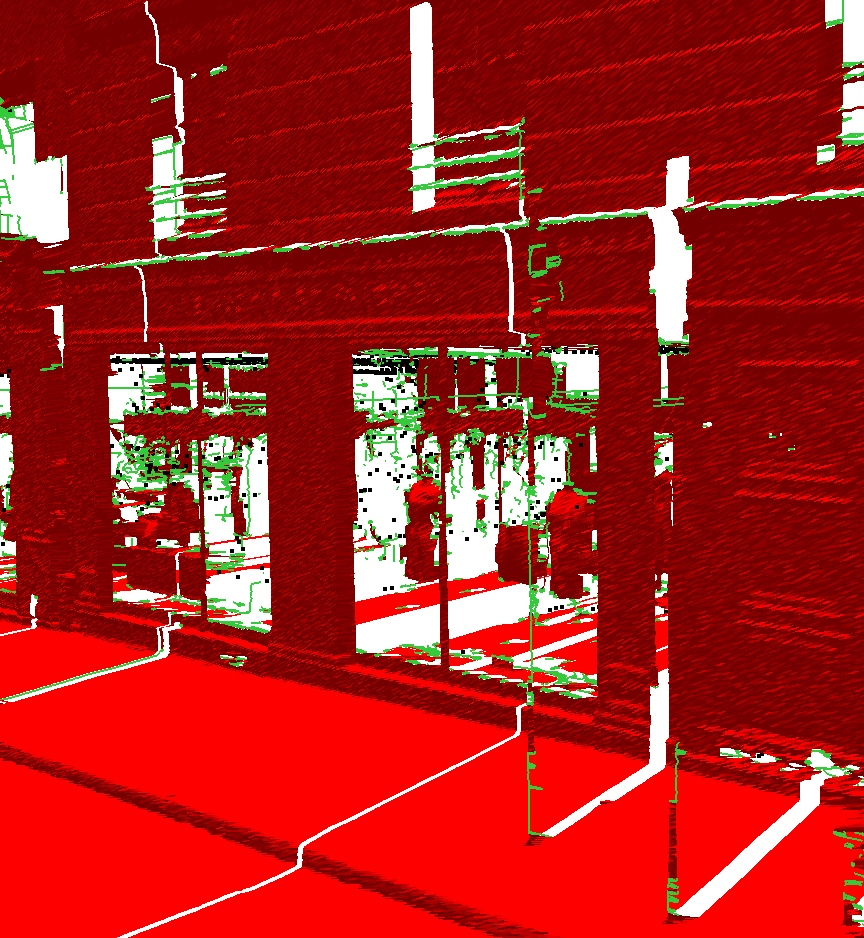}
\caption{Triangle filtering}
\end{subfigure}
\caption{Results on a complete urban scene, with road, facades, poles and pedestrians.}
\label{fig:urban}
\end{figure*}

Figure \ref{fig:urban} presents a reconstruction in a complex urban scene. Unlike the naive filtering, our methods are able to retrieve thin objects such as poles or windows bars without merging them to the closest objects. This is illustrated on figure \ref{fig:urban}. The top right image of this figure is an extract from Google Streetview to help the interpretation.

\begin{figure*}[t]
\centering
\begin{subfigure}[t]{1\textwidth}
\centering
\setlength{\tabcolsep}{3pt} % Default value: 6pt
\renewcommand{\arraystretch}{.7} % Default value: 1
\begin{tabular}{C{3.7cm}C{3.7cm}C{3.7cm}C{3.7cm}}%C{2.2cm}C{2.2cm}} 
%%%%% dont know why but the m{} command of package array won't work here
%Naive Filtering &
%\raisebox{2.5cm}[0pt][0pt]{Naive Filtering} &
\includegraphics[scale=.115]{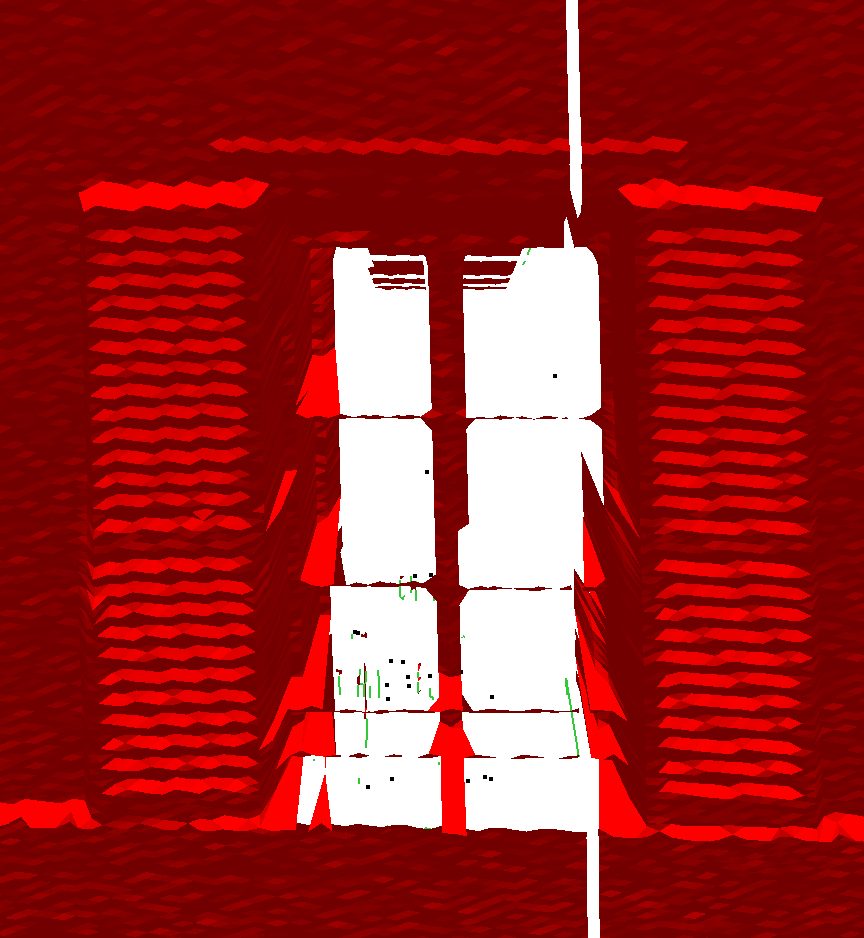} &
\includegraphics[scale=.115]{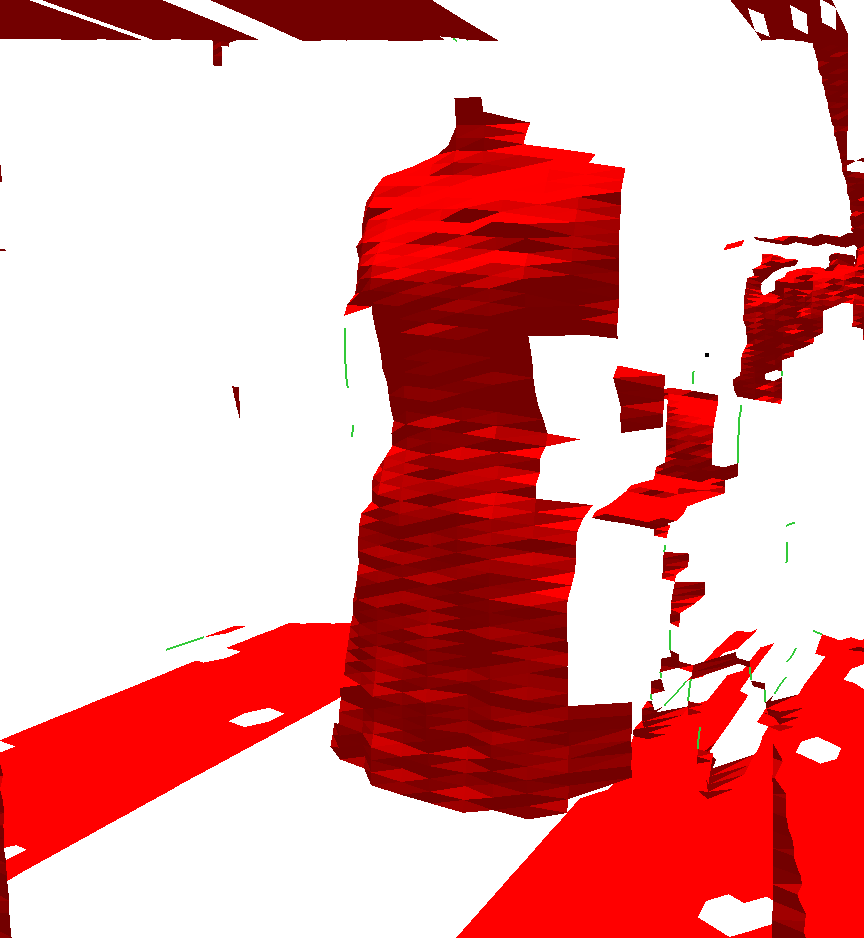} &
\includegraphics[scale=.115]{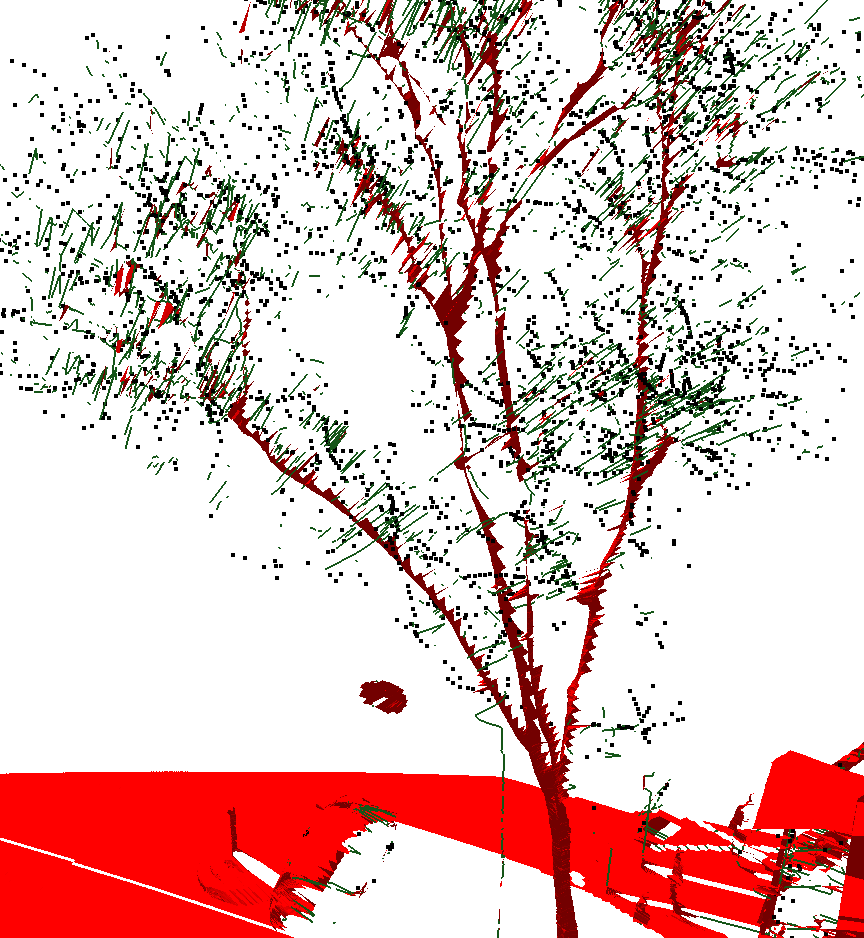} &
\includegraphics[scale=.115]{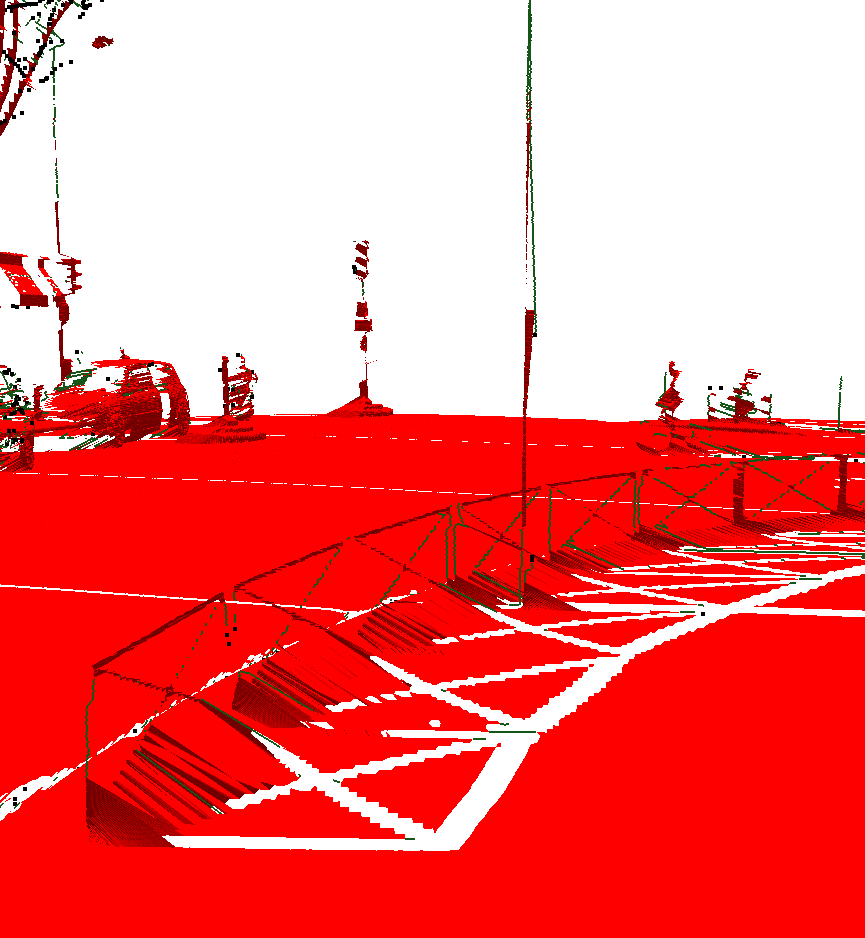}
\end{tabular}
\caption{Naive filtering}
\end{subfigure}

\begin{subfigure}[t]{1\textwidth}
\centering
\setlength{\tabcolsep}{3pt} % Default value: 6pt
\renewcommand{\arraystretch}{.7} % Default value: 1
\begin{tabular}{C{3.7cm}C{3.7cm}C{3.7cm}C{3.7cm}}%C{2.2cm}C{2.2cm}}
%Edge Filtering &
%\raisebox{2.5cm}[0pt][0pt]{Edge Filtering} &
\includegraphics[scale=.115]{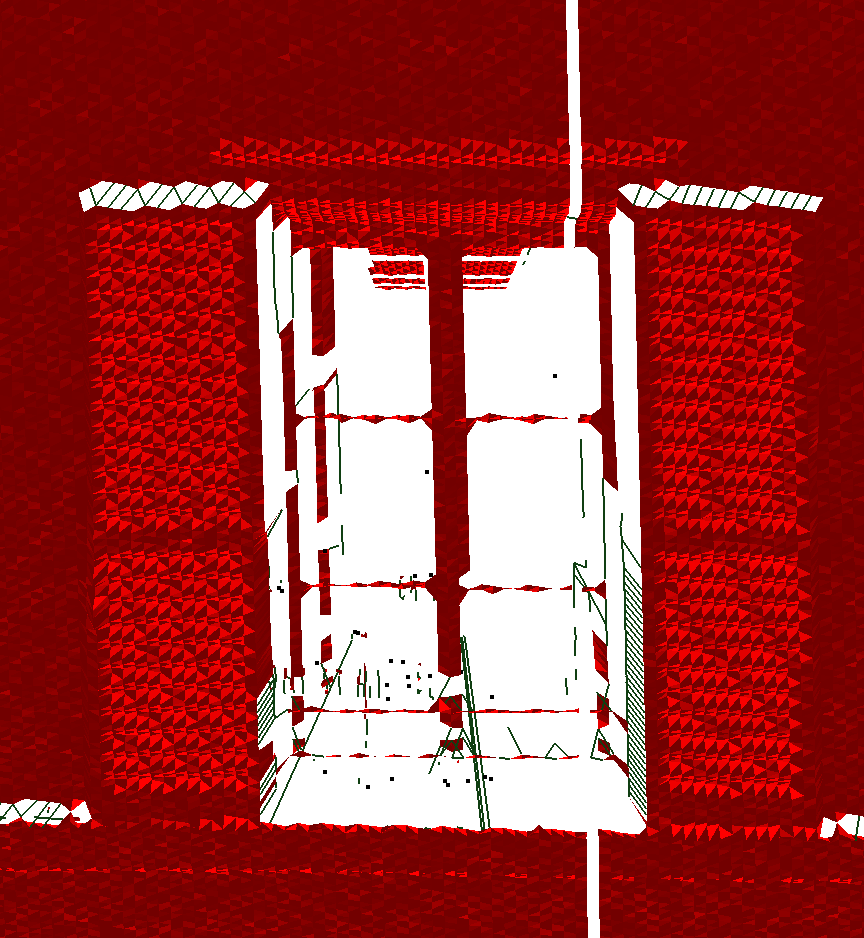} &
\includegraphics[scale=.115]{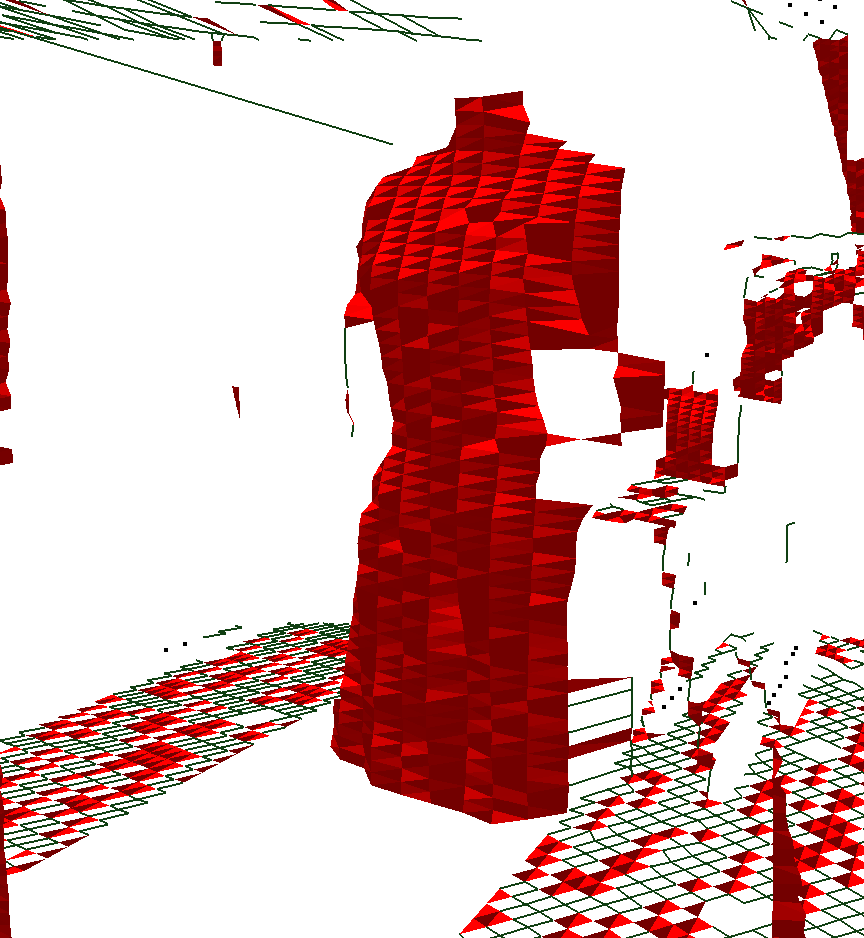} &
\includegraphics[scale=.115]{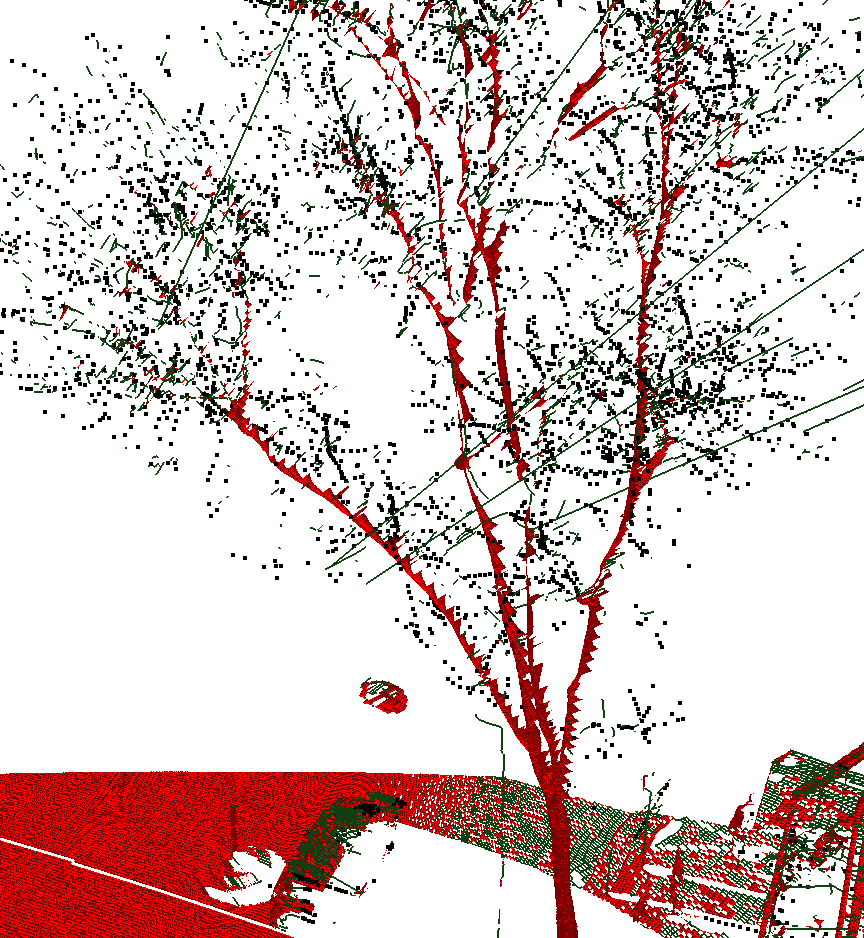} &
\includegraphics[scale=.115]{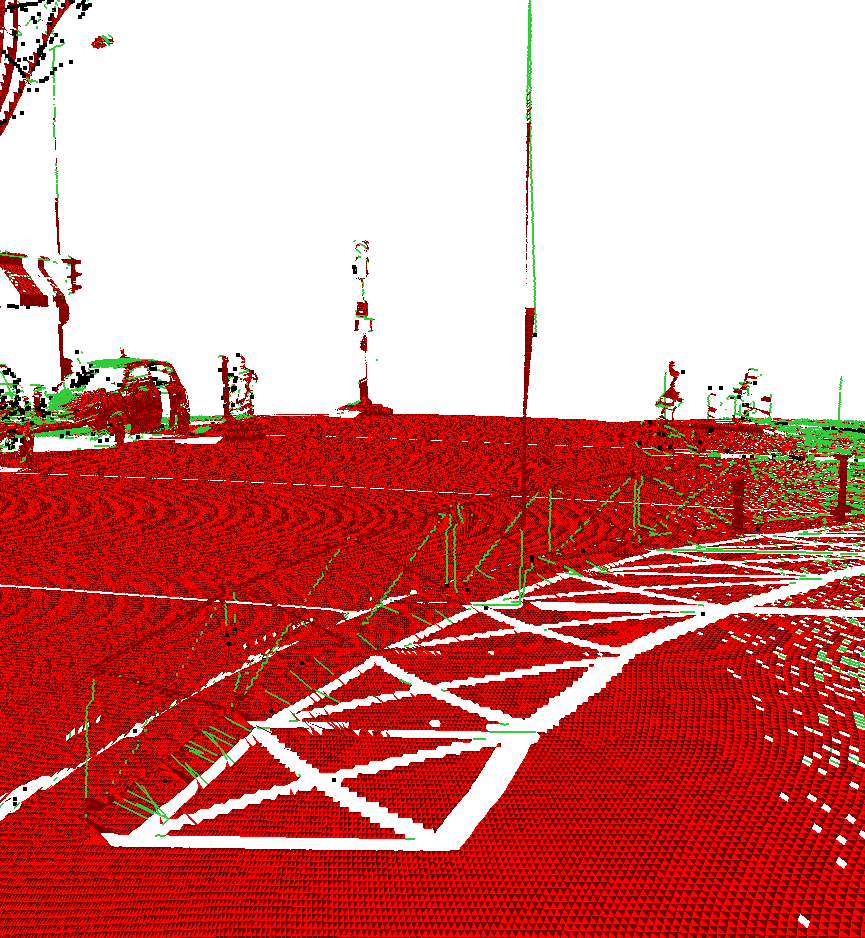}
\end{tabular}
\caption{Edge filtering}
\end{subfigure}

\begin{subfigure}[t]{1\textwidth}
\centering
\setlength{\tabcolsep}{3pt} % Default value: 6pt
\renewcommand{\arraystretch}{.7} % Default value: 1
\begin{tabular}{C{3.7cm}C{3.7cm}C{3.7cm}C{3.7cm}}%C{2.2cm}C{2.2cm}}
%Triangle Filtering &
%\raisebox{2.5cm}[0pt][0pt]{Triangle Filtering} &
\includegraphics[scale=.115]{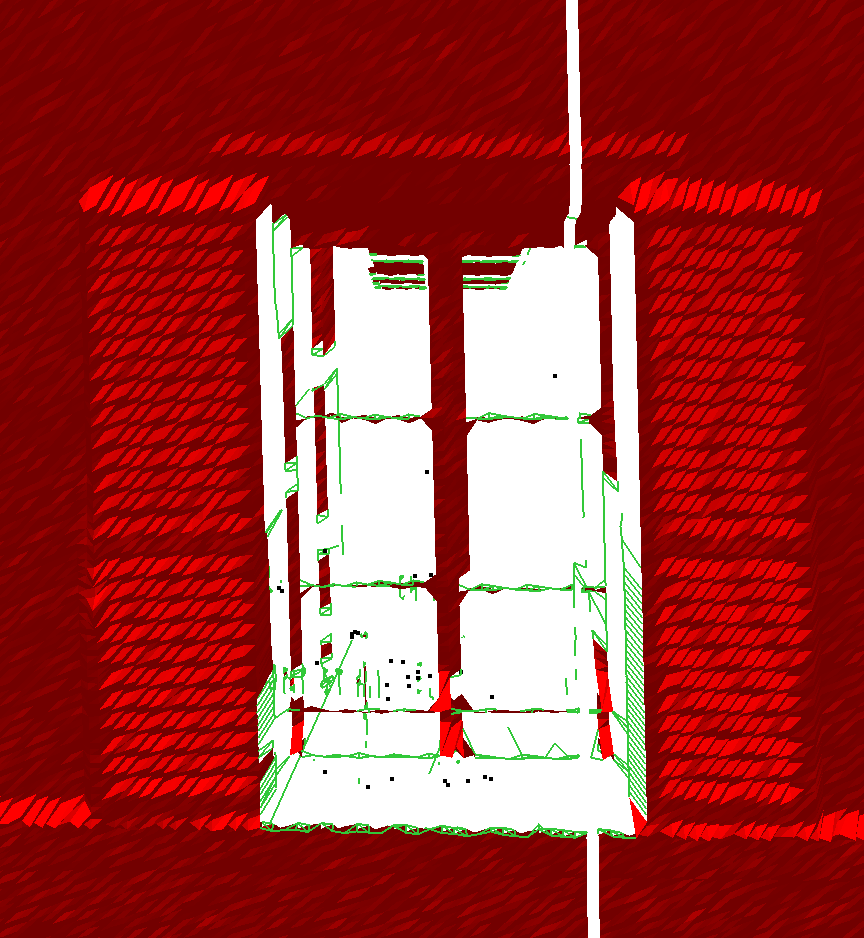} &
\includegraphics[scale=.115]{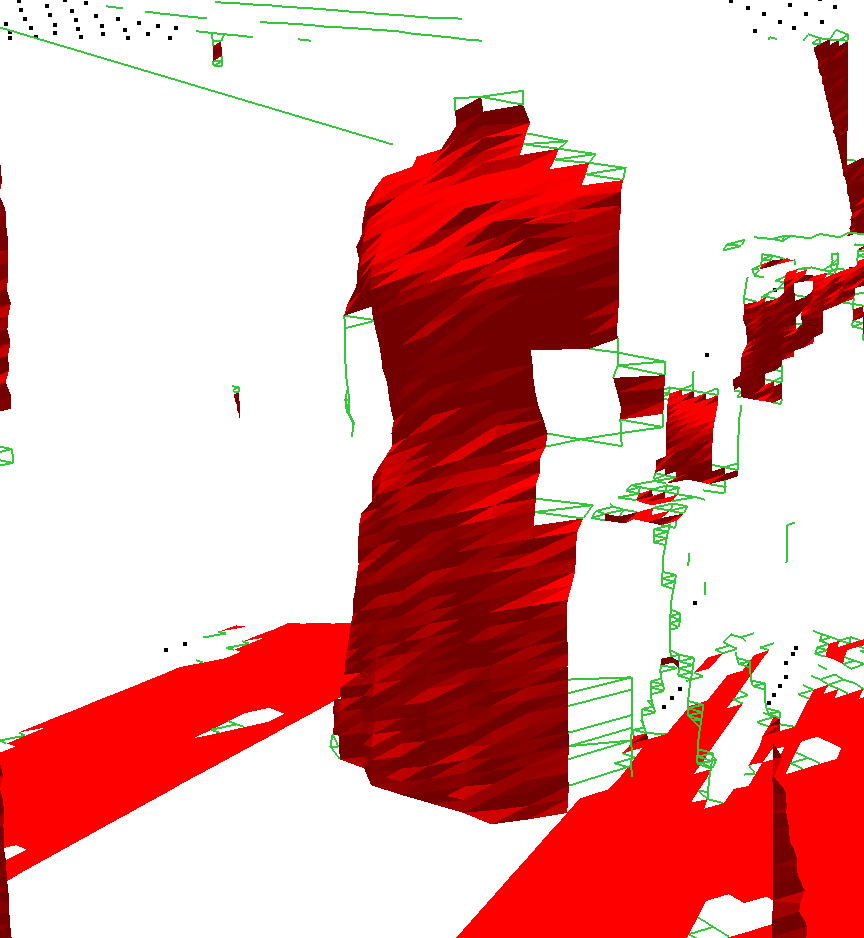} &
\includegraphics[scale=.115]{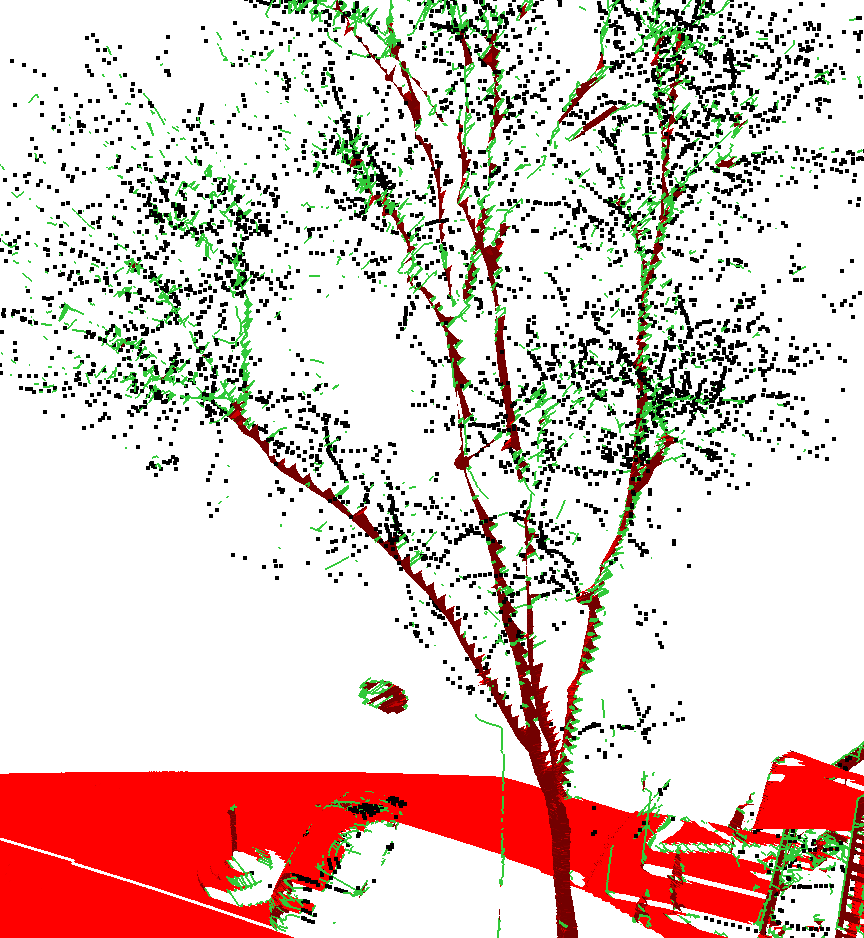} &
\includegraphics[scale=.115]{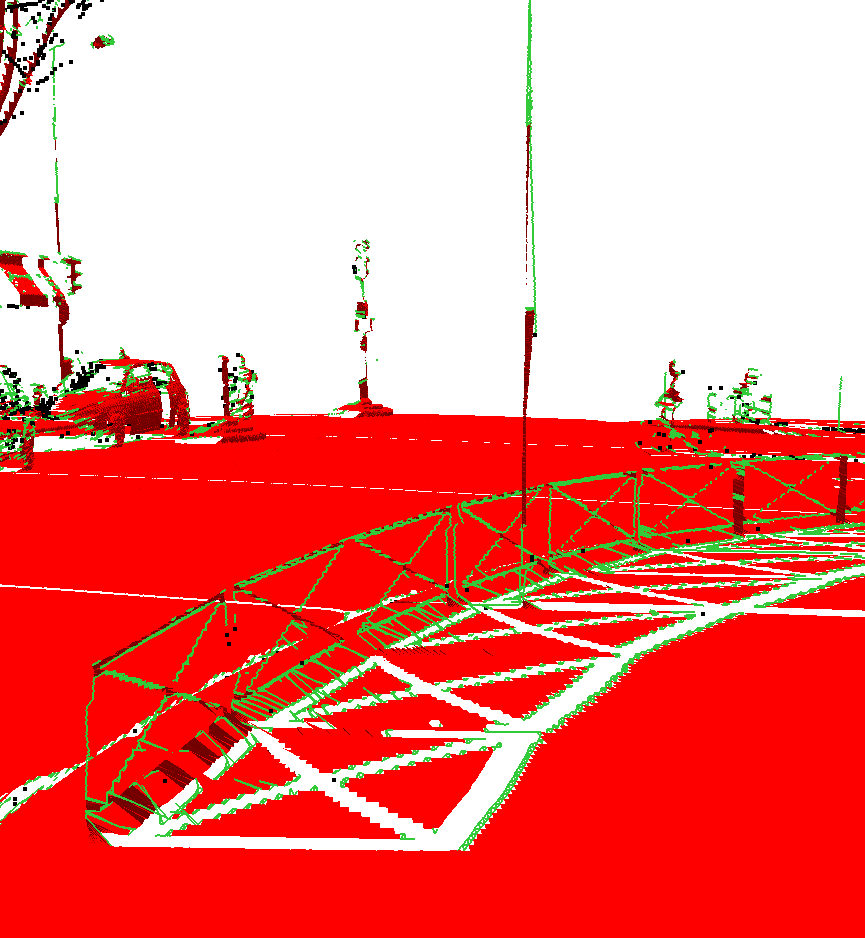}
\end{tabular}
\caption{Triangle Filtering}
\end{subfigure}
\caption{Comparison of the three methods: the naive filtering, the edge filtering and its extension with the triangle filtering. From left to right, the scenes represent: a window, a model in a showcase, a tree and barriers on a pavement.}
\label{fig:comparison}
\end{figure*}

Figure \ref{fig:comparison} focus more on specific areas of the scan. The first row shows the naive filtering method, whereas the second and third rows present respectively the edge filtering method and its extension with the triangle filtering. We remark that the naive filtering method struggles to retrieve limits between objects (like between poles and road, or people and buildings). The main advantage of the triangle filtering over the edge filtering that can be seen here, is that it helps to reduce the noise that occurs in complex areas such as tree foliage or grazing surfaces. We assume that in complex areas we cannot ensure the existence of connections between some points and will favor a reconstruction that remains careful on such areas. This is why the last method, which is less noisy than the edge filtering, is considered a more appropriate baseline for further developments, even if it discards some edges or triangles that had been well retrieved by the edge filtering method on grazing surfaces. 

% * <st.guinard@gmail.com> 2018-01-09T13:11:25.705Z:
% 
% J'ai retrouvé la scène sur maps (rue mabillon, près de l'angle avec la rue lobineau), et même si des éléments ont bougé depuis (piétons, vitrines) je me demandais si ça vaudrait le coup de la rajouter à côté pour montrer la réalité et que les reviewers comprennent mieux ce qu'ils voient
% 
% ^.

\section{Conclusions and perspectives}

This article presented a method for the simplicial complexes reconstruction of point clouds from MLS, based on the inherent structure of the MLS. We propose a filtering of edges possibly linking adjacent echoes by searching for collinear edges in the cloud, or edges perpendicular to the laser beams. We also presented an improvement of this method as a second filtering step, this time by looking for coplanar triangles. This last method produced simplicial complexes less holed than our first approach, and respects the noisy areas of the cloud (such as tree foliage) by discarding simplexes which existence cannot be ensured.

The main drawback of our methods is its high locality: we work only by considering point's neighbors and simplexes' adjacent simplexes. Using knowledge of the neighbors at different scales, or even on the whole cloud could help us to regularize the reconstruction according to more global structures of the cloud.
Further developments may also consider a hole filling process, to get rid of the absence of a few missing simplexes in a large structure (road, building). Last, setting up a generalization method, as in \citet{popovic1997progressive}, would be interesting to simplify the resulting simplicial complexes on large and regular structures in order to reduce the memory weight of the simplicial complexes while maintaining a high accuracy.

\section*{Acknowledgments}

The authors would like to acknowledge the DGA for their financial support of this work.

\section*{References}
\begingroup
\renewcommand{\section}[2]{}
\bibliography{Riva}
\endgroup
%\bibliography{Riva}

\end{document}